\documentclass[lettersize,journal]{IEEEtran}
\usepackage{amsmath,amsfonts}
\usepackage{algorithmic}
\usepackage{algorithm}
\usepackage{array}
\usepackage[caption=false,font=normalsize,labelfont=sf,textfont=sf]{subfig}
\usepackage{textcomp}
\usepackage{stfloats}
\usepackage{url}
\usepackage{verbatim}
\usepackage{graphicx}
\usepackage[inkscapelatex=false]{svg}
\usepackage{hyperref}
\usepackage{caption}
\usepackage{cite}
\usepackage{booktabs}
\usepackage{color}
\usepackage[flushleft]{threeparttable}
\usepackage{tablefootnote}
\usepackage{forest}
\usepackage{pdflscape}
\usepackage{afterpage}

\usetikzlibrary{shadows.blur}

\usepackage{booktabs,caption}

\begin{document}

\title{Fake News Detection through Graph-based Neural Networks: A Survey}

\author{
    Shuzhi Gong, Richard O. Sinnott, Jianzhong Qi \\
    The University of Melbourne, Melbourne, VIC 3000, Australia\\
    \texttt{\{shuzhi, rsinnott, jianzhong.qi\}@unimelb.edu.au} \\
    \and
    Cecile Paris \\
    Data61 CSIRO, Sydney, NSW 1710, Australia\\
    \texttt{Cecile.Paris@data61.csiro.au}
}




\maketitle

\begin{abstract}

The popularity of online social networks has enabled rapid dissemination of information. People now can share and consume information much more rapidly than ever before. However, low-quality and/or accidentally/deliberately fake information can also spread rapidly. This can lead to considerable and negative impacts on society. Identifying, labelling and debunking online misinformation as early as possible has become an increasingly urgent problem. 

Many methods have been proposed to detect fake news including many deep learning and graph-based approaches. In recent years, graph-based methods have yielded strong results, as they can closely model the social context and propagation process of online news.
In this paper, we present a systematic review of fake news detection studies based on graph-based and deep learning-based techniques. We classify existing graph-based methods into knowledge-driven methods, propagation-based methods, and heterogeneous social context-based methods, depending on how a graph structure is constructed to model news related information flows. We further discuss the challenges and open problems in graph-based fake news detection and identify future research directions. 

\end{abstract}


\section{Introduction}
\IEEEPARstart{O}nline social network platforms, such as Twitter and Reddit, offer immense convenience for users to share and consume content related to their daily lives. However, these platforms also facilitate the rapid, low-cost dissemination of rumors and/or fake news. Maliciously created fake news can have a significant negative impact on society, particularly during major events such as national  elections or pandemics. To combat fake news and reduce its detrimental effects, researchers have proposed various methods for automated fake news detection and classification. The advancement of deep learning techniques has ushered in a new era of fake news detection, with a large number of deep learning-based methods, particularly graph-based methods, being proposed. In this paper, we present a systematic review of such methods.

Given a news article comprising various content and contextual information, the task of fake news detection involves determining the veracity of the news and ideally automatically classifying it as either fake or real. Different approaches have analysed the news textual content and then expanded to gather information from  related entities. This includes not only explicitly related entities such as users who disseminate or comment on/reply to the news but also implicitly related entities, such as other news articles on  similar topics. 

The intricate, non-Euclidean relationships among such different entities related to a given news article can be represented using graph modeling techniques. For example, the BiGCN\cite{bian2020rumor} collects the all comments, retweets of a news item on Twitter platform and constructs a propagation graph to model the news spreading process, then uses a GCN to encode the graph and classify news item to fake or real one based on the graph representation.  
This has led to significant advancements in graph-based fake news detection in recent years. Graph modeling has been shown to offer immense potential in capturing non-Euclidean relationships between entities for fake news detection. However, despite the emergence and promise of graph-based methods, there has not yet been a systematic review of the background, progress, and developing trends in this area. This survey aims to fill this gap.


There are a number of existing surveys on fake news detection. However, the majority of works (e.g., \cite{shu2017fake, zhou2020survey, mridha2021comprehensive, hu2022deep}) focus on providing an overview of the entire field of fake news detection, and graph-based deep learning methods are either not mentioned at all ~\cite{shu2017fake} or only briefly described~\cite{zhou2020survey, mridha2021comprehensive, hu2022deep}. This lack of coverage creates a mismatch given the recent surge in the use of graph-based methods for fake news detection and the impressive results they achieve. One survey~\cite{varlamis-FutureInternet2022-GCNsurvey} exists on fake news detection using Graph Convolutional Networks (GCN), but it  only covers a few graph-based studies, since GCN is just one particular graph-based deep learning technique.

\begin{figure}
\centering
\includegraphics[width=0.48\textwidth]{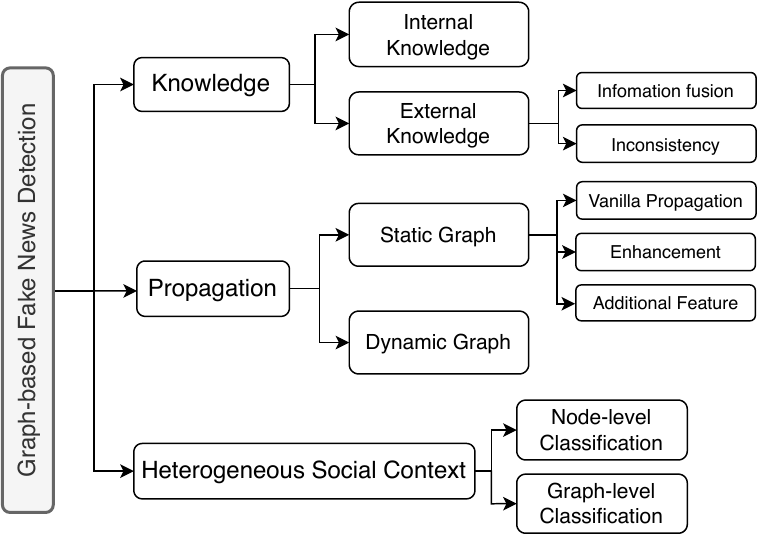}
\caption{A taxonomy on graph-based fake news detection methods.}
\label{Fig.taxonomy}
\end{figure}

\begin{figure*}[h]
\centering
\includegraphics[width=0.75\textwidth]{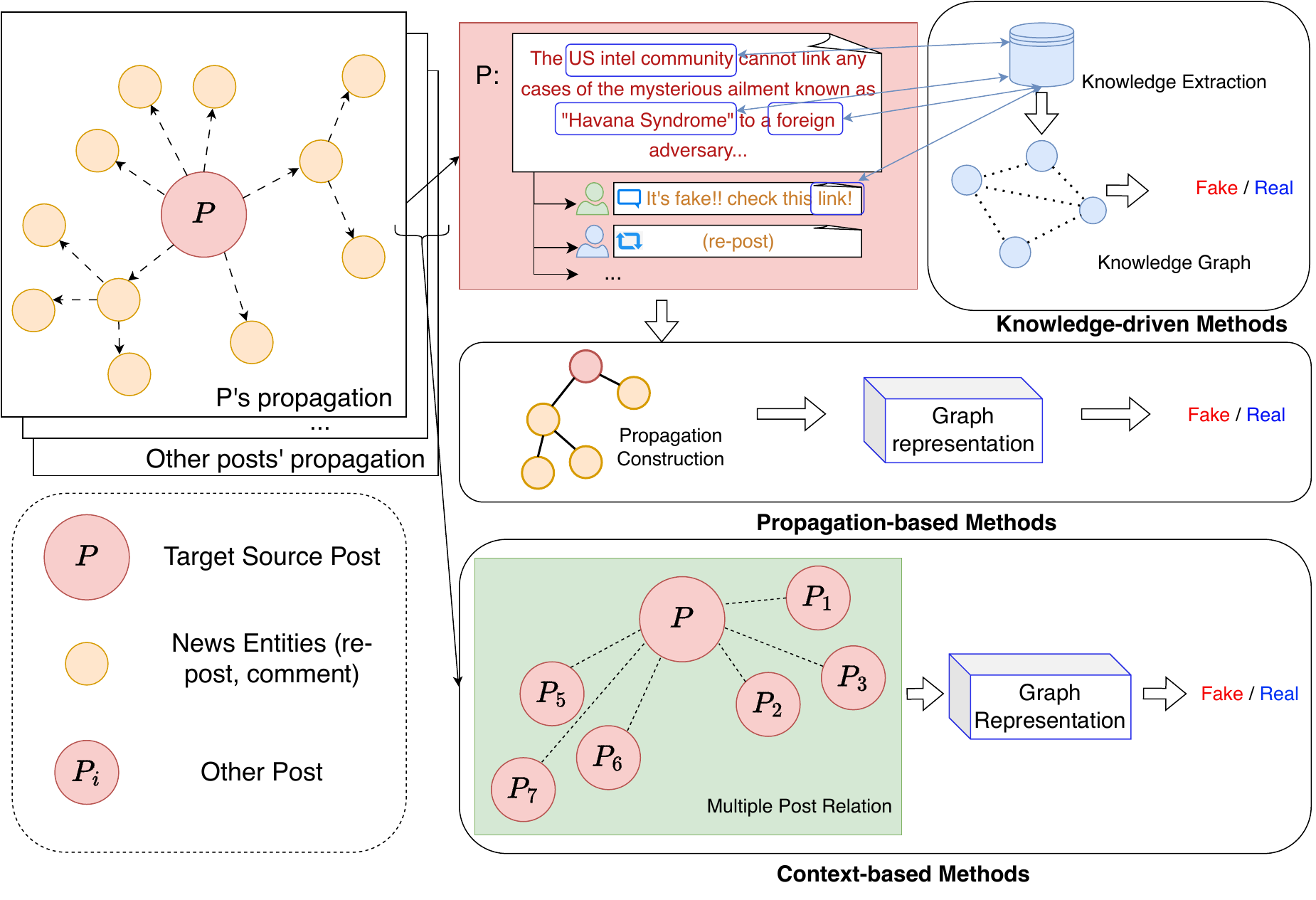}
\caption{Abstraction representation of knowledge-driven, propagation-based and social context-based methods.}
\label{fig_abstraction-of-3-methods}
\end{figure*}

Existing surveys typically provide a coarse-grained taxonomy of fake news detection studies, e.g., content- or social context-based approaches~\cite{shu2017fake}, whilst some methods take a hybrid approach considering both content and context. As such, a more fine-grained and nuanced taxonomy is needed. 

In this survey, we categorise graph-based fake news detection methods into three groups based on the underlying graph modelling methods: \emph{knowledge-driven methods}, \emph{propagation-based methods}, and \emph{heterogeneous social context-based methods} as illustrated in Fig.~\ref{Fig.taxonomy}.

Knowledge-driven methods leverage entities found in news content, e.g., concepts and named entities to identify fake news. Natural language processing (NLP) techniques can then be employed to pre-process the text content and extract such entities. Subsequently, a knowledge graph can be constructed to reflect the relationships among such entities. This can then serve as a computational representation of the news content. This graph is then encoded to graph embeddings using graph modeling techniques, and the embeddings are fed into a classifier to classify whether a given news article is fake  or real news. 
Sometimes,  the direct (internal) knowledge extracted from news contents may not be sufficient however, and hence external knowledge sources (e.g., Wikipedia) may need to be used. Entities in the news content can be linked to such external sources through entity linking~\cite{shen2014entity} to obtain  more comprehensive information. Both internal and external knowledge can then be combined for fake news detection. As a result, knowledge-driven methods can be further categorised into internal and external knowledge-based methods.

Propagation-based methods focus on the dissemination process of news articles. Throughout the process, many users typically interact by posting, reposting or replying to a given article. These users and their interactions form a tree (or graph)  structure. By examining the news propagation structure and the trustworthiness of the users within the propagation network, the potential veracity of a given news article can be inferred.

Heterogeneous social context-based methods extract the context of the source news, such as the other posts that come from the same user together with other news items on the same topic. Such a social context can also be represented with a graph that helps the fake news classification process.

One key difference between propagation-based methods and heterogeneous social context-based methods is the process whereby graph modelling is applied. When a graph is used  to model the propagation process of individual news items, it implies a propagation-based method. 
When a graph is applied to model a larger social context involving multiple news articles and users interacting on different or related news articles, it requires that a heterogeneous social context-based method is adopted. In this case, the veracity of multiple news articles needs to be inferred from the graph. 

Fig.~\ref{fig_abstraction-of-3-methods} visualises and summarises the differences among the three types of methods. 


Overall, the contributions of this survey are as follows: 
\begin{itemize}

\item \textbf{A systematic review.} We provide a systematic review of existing graph-based fake news detection methods, describing how they were developed and discuss their strengths and weaknesses. 

\item \textbf{A novel taxonomy}. We propose a novel taxonomy focusing on graph-based deep learning methods for fake news detection. 

\item \textbf{Future direction landscape.} We discuss open problems and challenges in  graph-based fake news detection methods, providing insights on future research directions. 


\end{itemize}

The rest of the paper is organised as follows. In Section~\ref{sec.preliminaries}, we define the core concepts related to fake news detection and provide a definition of the three categories of fake news detection methods. Next, an overview of the papers reviewed is given in Section~\ref{sec.overview}. We describe representative methods of the three categories in Sections~\ref{sec.knowledge-driven},~\ref{Sec.propagation-based-methods} and~\ref{sec.heterogeneous-context-based-methods}. Open issues and opportunities are discussed in Section~\ref{sec.challenges-and-opportunities}, and widely used datasets are presented in Section~\ref{sec.dataset}.

\section{Preliminaries}
\label{sec.preliminaries}
We start by some basic concepts and definitions related to graph-based fake news detection. \emph{Fake news} is a news article that is intentionally or verifiably shown to be false~\cite{shu2017fake}. This  definition entails two characteristics of fake news: fake news can be verified to be false, and they often come with a dishonest intention to mislead readers. \emph{Fake news detection} is a process used to detect fake news items, e.g., Twitter posts. As noted, in this survey we explore graph-based fake news detection based on the following methods: 

\begin{itemize}
\item \textbf{Knowledge-driven methods} construct a knowledge graph for a given news article $\mathbf{a}$ which is then used to establish the article veracity. The knowledge graph $\mathbf{K_{g}}$ is typically constructed from entities $\mathcal{En} = \{ en_{i}\}$ extracted from the textual and potentially visual content of the news article. In the external knowledge-based methods, the $ \mathcal{En} $ also includes entities from external knowledge database. 

\item \textbf{Propagation-based methods} model the propagation process of a given news item by constructing an associated \emph{propagation graph}. This graph is then used to assess the article veracity. The propagation graph of a given news item 
$\mathbf{a}$ is formed by a set of engagement tuples $\mathcal{E} = \{ e_{it}\}$ that represents the process of how $\mathbf{a}$ spreads over time $t$ among $n$ users  $\mathcal{U} = \{ u_1, u_2, \ldots, u_n\}$, and their corresponding posts $\mathcal{P} = \{ p_1, p_2, \ldots, p_n\}$ and re-posts/comments. Each engagement $e_{it} = \langle u_i, p_i, t\rangle$ represents a user $u_i$ resharing (spreading) news article $\mathbf{a}$ at time $t$ and optionally with a  comment $p_i$. 

\item \textbf{Heterogeneous social context-based methods} consider the broader context of news items for fake news detection. A graph covering multiple news items $\mathbf{A} = \{ \mathbf{a_1}, \mathbf{a_2}, \ldots, \mathbf{a_1}\}$ and their engagements is considered, e.g., all news items from the \emph{same user} who posted the news item being evaluated for fake news, or news articles from other users but on the \emph{same topic}. Since the social context  is formed by different types of entities (e.g., users and news articles) and connections (e.g, user-user follower/followee connections, user-news post connections), a heterogeneous social context graph is constructed by the various methods in this category.

\end{itemize}

We note a related concept of \emph{rumour}. A rumour is unverified information, which may not necessarily be real or fake. When a rumour is identified as false,  the rumour is regarded as fake news.  Many studies often use the terms ``rumour'' and ``fake news'' interchangeably. In this survey, we also cover graph-based rumour detection methods.  

\begin{table*}
    
\setlength{\columnsep}{0pt}
\centering
  \captionsetup{justification=centering}
  \caption{Summary of Graph-based Fake News Detection Methods.}
  \label{tab:commands}
\resizebox{\textwidth}{!}{
  \begin{threeparttable}
  \begin{tabular}{p{2.6cm} p{3cm} p{3cm} p{2.0cm} p{6cm}}
    \toprule
    \textbf{Models} & \textbf{Graph Nodes} & \textbf{Edge Construction}  & \textbf{Graph Encoder} & \textbf{Core Strategy}\\
    \midrule
    \multicolumn{5}{c}{\textbf{Knowledge-driven Methods}}\\
    \midrule
    \textbf{GCN-text}\cite{vaibhav2019sentence} & sentences & fully-connected & GCN \cite{kipf2017semisupervised-gcn} & Capture sentence connections in article. \\
    \textbf{GET}\cite{Xu-WWW22-GET} & words & sliding window \cite{yao2019graph} & GGNN \cite{Xu-WWW22-GET} & Capture long-distance semantic dependency. \\
    
    \textbf{FinerFact}\cite{jin2022towards} & claims, posts, users, keywords & fully-connected, propagation & KGAT \cite{liu-acl2020-fine} & Extract claims from news content and evidence from comments. \\
    \textbf{KMGCN}\cite{wang2020KMGCN} & visual, textual, external knowledge entities & PMI \cite{church1990word-pmi} & GCN &  Understand the news content through external knowledge.    \\
    \textbf{KMAGCN}\cite{qian2021KMAGCN} &  textual and external knowledge entities & PMI & Adapt-GCN \cite{qian2021KMAGCN} & Understand the news through external knowledge like KMGCN.\\
    \textbf{LOSIRD}\cite{li2021meet} & comments, retrieved evidence & linking evidence to source post, propagation & GraphSAGE \cite{hamilton2017inductiveGraphSAGE} & Retrieve external evidence for news source post. \\
    \textbf{CompareNet}\cite{hu-etal-2021-compare} & topics, knowledge entities, sentences & TransE\cite{bordes2013translating} & Hetero-GCN\cite{hu-etal-2021-compare} & Compare the news claims with external knowledge.  \\
    \textbf{InforSurgeon}\cite{fung2021infosurgeon} & visual, textual, external knowledge entities & subject-predicate-object & neural network & Capture news content and external knowledge inconsistency. \\
    \midrule
    \multicolumn{5}{c}{\textbf{Propagation-based Methods}} \\ 
    \midrule
    \textbf{RvNN}\cite{ma2018rumor} & claim, comments & responsive propagation & - \tnote{*} & RNN encodes posts in tree-shaped order.\\
    \textbf{Tree-Transformer}\cite{ma2020debunking} & claim, comments & responsive propagation & -  & Transformer encodes posts in tree-shaped order. \\
    \textbf{TRM-CPM}\cite{li2020exploiting} & claim, comments & responsive propagation & GraphSAGE & Model the responsive propagation by a graph. \\
    \textbf{Bi-GCN}\cite{bian2020rumor} & claim, comments & responsive propagation & GCN & Model rumour spreading and diffusion in a graph from both top-down and bottom-up directions. \\
    \textbf{EBGCN}\cite{wei-etal-2021-towards} & claim, comments & responsive propagation (with tuning)& GCN & Infer edge weights from node features by the Bayesian theorem. \\
    \textbf{UPSR}\cite{wei2022UPSR} & claim, comments & responsive propagation (with tuning) & GCN & Enhance edge connections by Gaussian distributions and KL-divergence. \\
    \textbf{EDEA}\cite{SIGIR2021-RDEA}, \textbf{GACL}\cite{sun2022rumor}, \textbf{RDCL}\cite{ma2022towards}, \textbf{CCFD}\cite{ma-CIKM2022-curriculum} & claim, comments & responsive propagation (with tuning)& GCN & Introduce adversarial and contrastive learning when modeling the propagation 
    graph. \\
    \textbf{GCAN}\cite{lu-li-2020-gcan} & users & fully-connected & GCN & Model potential interaction among users. \\  
    \textbf{UPFD}\cite{dou2021user} & claim, responsing users' historical posts & responsive propagation & GNN & Retrieve user historical posts as user preference. \\
    \textbf{PSIN}\cite{min2022divide} & claim, users, comments & responsive propagation, user-post authorship, user-user following & GAT\cite{velivckovic2017graph} & Consider multiple types and relations in the news propagation process. \\
    \textbf{DUCK}\cite{tian2022duck} & claim, comments, retweets, users & responsive propagation & GAT& Consider temporal and structural information of comments and retweets in propagation. \\
    \textbf{UniPF}\cite{wei-coling2022-UniPF} & claims, comments, topics & cluster connection, responsive propagation & Adapt-GCN\cite{wei-coling2022-UniPF} & Connect claims under the same topic.  \\
        \textbf{RNLNP}\cite{lao2021rumor} & claim, comments & responsive propagation & GCN & Introduce linear propagation information in addition to non-linear graph modelling. \\
    \textbf{Dynamic-GCN/GNN}\cite{choi2021dynamic,song2022dynamic} & claim, comments & responsive propagation & GCN & Take snapshots of the propagation graph to capture temporal information. \\
    \textbf{TGNF}\cite{song2021temporally} & claim, comments & responsive propagation & TGN \cite{xu2020inductive} & Use advanced temporal graph model to capture temporal information of the propagation graph evolution process. \\

    \textbf{DDGCN}\cite{Sun_Zhang_Zheng_Ma_2022} & claim, comments, knowledge entities & responsive propagation, knowledge graph & GCN & Take snapshots of both the propagation and the knowledge graphs. \\
    \textbf{SEAGEN}\cite{gong2023fake} & claim, comments, retweets  & responsive propagation & TGN, hawkes process & Capture the self-exciting graph evolution process.  \\
    \midrule
    \multicolumn{5}{c}{\textbf{heterogeneous social context-based Methods}} \\ 
    \midrule
    \textbf{GLAN}\cite{yuan2019GLAN} & news, users, retweets & news-retweet-user & CNN & Jointly learn the local (in-news) and global (cross-news) relations. \\
    \textbf{Monti}\cite{monti2019fake} & news, users & user-user following, retweeting propagation & GCN & Model news propagation with global user following relationships. \\
    \textbf{NDG}\cite{kang2021fake} & news, users, sources, comments, domains & news-domain, source-news authorship, responsive propagation & HDGCN\cite{kang2021fake} & Introduce more social context information. \\
    \textbf{MFAN}\cite{mfan2022zheng} & news, users, comment posts & authorship, responsive propagation & sign GAT \cite{mfan2022zheng} & Capture implicit social connections.\\
    \textbf{SureFact}\cite{yang-kdd2022-reinforcement} & claim, users, comments, retweet, keywords &authorship, responsive propagation & KGAT & Capture informative parts of the propagation graph. \\
    
    \textbf{TriFN}\cite{shu2019beyond} & publishers, news, spreaders & publishers-news-spreaders connection & - & Exploit social context.  \\
    \textbf{FANG}\cite{nguyen2020fang} & news sources, news, users & user-user following, source-source citation, source-news publication, etc. & GraphSAGE & Exploit social context. \\
    \textbf{Mehta}\cite{mehta2022tackling} & news sources, news, users & FANG-like connections, link inference & R-GNN\cite{schlichtkrull2018modeling} & Inference implicit connections in social context. \\ 
    \textbf{Hetero-SCAN}\cite{cui-CIKM2022-HeteroSCAN} & publishers, news, spreaders & publisher-news, news-spreader relations & - & Capture multi-level and temporal information in social context.\\
    \textbf{TR-HGAN}\cite{gao2022topology}  & news,  comments, users & authorship, responsive propagation &  hierarchical attention~\cite{gao2022topology} & Mitigate topology imbalance and relation in-authenticity in the heterogeneous social context. \\
    \bottomrule
  \end{tabular}
  \begin{tablenotes}
  \item[*]: Does not use graph modelling techniques in the encoding process. 

  \end{tablenotes}
  \end{threeparttable}
  }
  \label{Tab.methods_overview}
\end{table*}

\section{Overview of Graph-based Fake News Detection Methods}
\label{sec.overview}

Table~\ref{Tab.methods_overview} summarises key studies using graph-based methods for fake news detection. It outlines how the graph structures are constructed and encoded. 

\textbf{Graph construction.} To form graph nodes, knowledge-driven methods tend to use sentences, words, topics or knowledge entities from target news articles, while most propagation-based methods use the news articles and associated comments. While  news articles and comments also contain words and sentences, the focus of the propagation-based methods is more on their  propagation pattern instead of the detailed news contents. Heterogeneous social context-based methods also consider the propagation patterns but they emphasise the connections between multiple news items. Thus, besides the news and comments, users and topics (or keywords) also form graph nodes that serve as bridges between the graph nodes.

To connect nodes in the graph (i.e., to create the edges), a common strategy used by knowledge-driven methods is to simply connect all pairs of nodes. The PMI~\cite{church1990word-pmi} strategy is a variant, which forms a fully connected graph where the weight of an edge is the similarity between two nodes, e.g., textual similarity. Propagation-based methods, on the other hand, create edges based on propagation events, e.g., adding an edge to connect a comment node with the article node being commented on. Heterogeneous social context-based methods aim to capture every connection in the social context. Edges are added among users, news, comments and even topics. 

\textbf{Graph encoding.} 
After the graph structures have been constructed, they 
are typically encoded with a deep learning module, e.g., a \emph{graph encoder}  captures latent information of the graphs for fake news classification.  
 Graph Convolutional Networks (GCN) are a common choice for  graph encoders and used by 
more than half of the methods surveyed. Several variants of GCN have been proposed~\cite{Xu-WWW22-GET,qian2021KMAGCN,hu-etal-2021-compare,wei-coling2022-UniPF,kang2021fake}. Besides GCN, Graph Attention Networks (GAT) \cite{velivckovic2017graph} and its variants (e.g., KGAT~\cite{liu-acl2020-fine} and TGN~\cite{xu2020inductive}) have been put forward based on their strong capability for identifying important neighbour nodes in a graph, e.g., nodes that include more important words related to a given news article. 


\textbf{Fake news detection strategies.} 
As Fig.~\ref{Fig.taxonomy} shows, knowledge-driven methods can be further categorised into \emph{internal knowledge-driven} ~\cite{vaibhav2019sentence,Xu-WWW22-GET,jin2022towards} and \emph{external knowledge-driven}~\cite{hu-etal-2021-compare,fung2021infosurgeon} methods. Internal knowledge-driven methods 
utilise a graph to model the semantic relationships within the news content and social context. This works in a way similar to some text classification methods~\cite{liu2023document}. In this model, the news article including comments from other users is regarded as a text document. Then, connections between different parts of the document are extracted and modelled via a graph, and fake news detection is performed as a graph classification task.  External knowledge-driven methods utilise knowledge from external resources. References to and comparisons with external authoritative knowledge sources are enabled. 

For propagation-based methods, the strategy is to model  responsive propagation patterns (i.e., how one user post responds to another) and classify news based on the patterns. Considering that the patterns can contain noise and be modified by fake news spreaders, edge enhancement, adversarial learning, and contrastive learning strategies are often used~\cite{wei2022UPSR,SIGIR2021-RDEA,sun2022rumor,ma2022towards,ma-CIKM2022-curriculum}. There is also a trend to combine more information with the propagation patterns, such as  temporal information~\cite{lao2021rumor,choi2021dynamic,song2022dynamic,song2021temporally,Sun_Zhang_Zheng_Ma_2022,gong2023fake} and user information~\cite{dou2021user,min2022divide,tian2022duck}.

Heterogeneous social context-based methods can be seen as a zoomed-out version of propagation-based methods. They use a heterogeneous graph to model the global social context  including connections between different news articles. 

In recent years, more and more hybrid models have been proposed which fall into more than one of the aforementioned method categories. For example,  DDGCN~\cite{Sun_Zhang_Zheng_Ma_2022} is a propagation-based method that takes snapshots of the propagation graph and the knowledge graph. Similarly, some heterogeneous social context-based methods~\cite{kang2021fake,mfan2022zheng} also follow the idea of propagation-based methods, to model propagation patterns by GNNs, but with additional information obtained from a larger social context. 

In the next three sections, we detail the studies in each of the three categories.

\section{Knowledge-driven methods}
\label{sec.knowledge-driven}

\begin{figure}[h]
\centering
\includegraphics[width=0.49\textwidth]{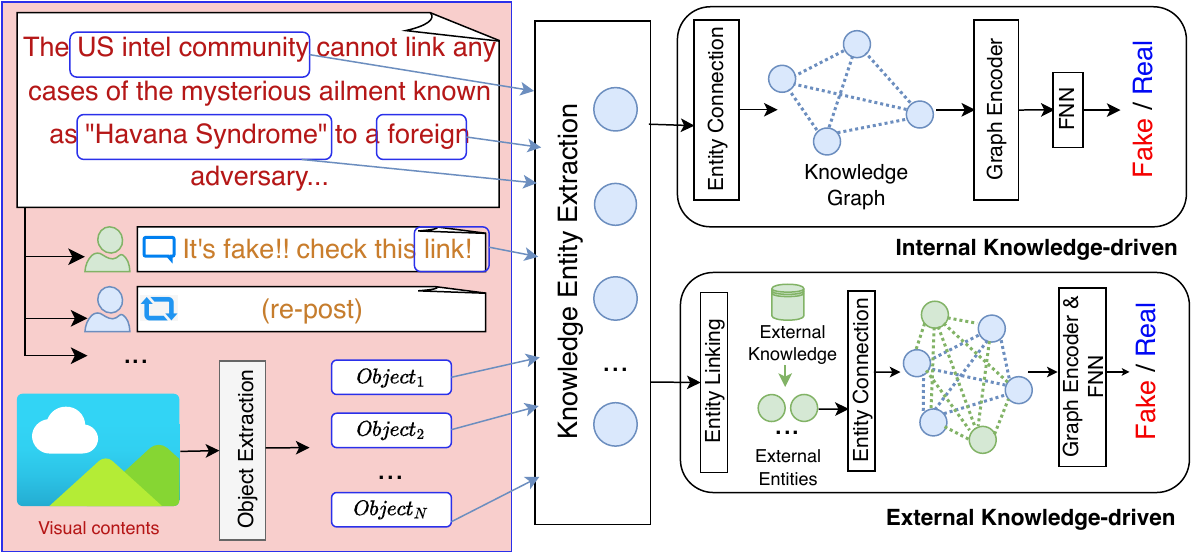}
\caption{Basic idea of internal and external knowledge-driven methods.}
\label{fig_abstraction_knowledge_driven}
\end{figure}


Knowledge-driven fake news detection methods, as shown in Fig.~\ref{fig_abstraction_knowledge_driven}, extract entities (e.g., nouns or named entities) from the news content and sometimes  from reader comments to construct a (knowledge) graph. The graph describes the connections (relationships) among the entities. In most cases, the nodes in the graph represent the  entities, while the edges represent the connections. Fake news detection is then done by analysing and performing anomaly detection on the knowledge graph constructed. 

As mentioned above, depending on whether external knowledge sources (e.g., YAGO~\cite{suchanek2007yago} or Probase~\cite{song2011probase}) are used, knowledge-driven methods can be further divided into internal knowledge-based or external knowledge-based approaches. Different from internal knowledge-driven methods, external knowledge-driven methods collect external knowledge entities from external databases based on the news contents, using entity linking techniques~\cite{shen2014entity} (cf. Fig.~\ref{fig_abstraction_knowledge_driven}).  
As a result, graphs constructed through external knowledge-based methods contain both entities that are internal and external to the news content, while those constructed by internal knowledge-based methods only contain internal entities.

The graphs constructed by  knowledge-driven methods are 
processed by graph modelling techniques (i.e., graph encoders). The graph encoder outputs are usually fed into fully connected neural networks to perform binary classification, i.e., to classify whether the graph corresponds to fake or real news. Following such a pipeline, existing methods mainly differ in how the knowledge graphs are constructed and encoded. 


\subsection{Internal Knowledge-Based Methods}

Internal knowledge-based methods construct a knowledge graph to model the semantics of the contents of a news item. They use graph encoders to encode the knowledge graphs  and perform graph classification for fake news detection. Studies following this approach mainly vary in the knowledge graph construction process, including the knowledge entity selection (i.e., using sentences, words, or claims from the news content) and/or the knowledge entity connection (fully-connected or sliding-window based connections). 

The internal knowledge-based models are trained in a supervised manner, by propagating the loss calculated based on the predicted news veracity and the ground-truth veracity. Cross-entropy loss is often used as the basis for the loss function: 
\begin{equation}
\mathcal{L} = -y\log(\hat{y}) + (1-y)\log(1-\hat{y}) + \lambda_{reg} ||\Theta||^{2}
\label{Eq.cross-entropy}
\end{equation} 
where $y$ is the ground-truth news veracity, $\hat{y}$ is the prediction result, $\lambda_{reg}$ is the regularisation coefficient, and $||\Theta||^2$ is the $L_2$ norm of the model parameters, i.e., $\lambda_{reg} ||\Theta||^{2}$ is the $L_2$ regularisation. GCN-text~\cite{vaibhav2019sentence}, GET~\cite{Xu-WWW22-GET}, DISCO~\cite{fu-cikm2022-disco} all utilise this standard cross-entropy loss, i.e., setting $\lambda_{reg}$ to $0$ in Equation~\ref{Eq.cross-entropy}, while  FinerFACT~\cite{jin2022towards} uses a non-zero value for $\lambda_{reg}$. We discuss these methods below.

\textbf{GCN-text}~\cite{vaibhav2019sentence} considers the differences in the context around sentences associated with real news and fake news. It builds a graph where every sentence of a news item is represented as a node, and the sentence interactions are represented by edges. The  features of a node are the  textual representation of the sentence computed by a \emph{Long Short-term Memory} (LSTM) network. To consider all possible interactions, the graph is constructed to be fully connected. Then, the fake news detection task is transformed into a graph classification task. The news graph is encoded by a GCN which corresponds to Equation~\ref{GCN_formula}:

\begin{equation}
\label{GCN_formula}
\mathbf{H}^{(l+1)} = \sigma(\Tilde{\mathbf{D}}^{-\frac{1}{2}}\Tilde{\mathbf{A}}\Tilde{\mathbf{D}} \mathbf{H}^{(l)} \mathbf{W}^{(l)} )
\end{equation}
Here, $\mathbf{H}^{(l)} \in \mathbb{R}^{N \times D}$ is the computed node feature matrix at layer $l$, for the $N$ nodes in the input graph and $D$ features per node where $\mathbf{H}^{(0)} = \mathbf{X}$ is the input node feature matrix; $\mathbf{W}^{(l)} \in \mathbb{R}^{D \times D'}$ is the weight matrix at layer $l$, with $D'$ output features per node; $\tilde{\mathbf{A}} = \mathbf{A} + \mathbf{I}_N$ is the adjacency matrix of the graph with added self-loops; $\tilde{\mathbf{D}}$ is the diagonal degree matrix of $\tilde{\mathbf{A}}$; $\sigma(\cdot)$ is the activation function, applied element-wise to the resultant matrix, and $\mathbf{H}^{(l+1)} \in \mathbb{R}^{N \times D'}$ is the output feature matrix at layer $l+1$ with $D'$ features per node.

The output graph embedding is processed by a pooling layer, and the pooled representation is then utilised for fake news classification using fully connected layers.

\textbf{GET}~\cite{Xu-WWW22-GET} constructs knowledge graphs where every single word of a news article is represented as a node. Edges are built between words (i.e., nodes) within fixed-size sliding windows. In the knowledge entity extraction step (cf.~Fig.~\ref{fig_abstraction_knowledge_driven}), entities are from both the \emph{source news} (i.e., the news article to be classified) and the comments. Several word-based knowledge graphs are constructed for each source news and associated comment. Then, GET follows the same procedure as GCN-text based on applying GCN on the word-based knowledge graphs, feeding the graph representation into fully connected layers with \textit{attention mechanism}, and classifying news to be fake or real. With the assistance of the attention mechanism, GET is able to tell which words provide fake information (i.e., the \emph{misleading words}).
In a similar fashion, \textbf{DISCO} \cite{fu-cikm2022-disco} also constructs a graph to model word relationships in news content. It only extracts knowledge entities from source news, and hence only one knowledge graph is constructed for each news article. DISCO reveals misleading words by masking different nodes in the knowledge graph to help observe their contributions to the classification outcome. 


By building and analysing knowledge graphs over news content, knowledge-driven methods can provide explanations about why a news article is classified as fake news, thus enabling result interpretability. GET\cite{Xu-WWW22-GET} and DISCO\cite{fu-cikm2022-disco} offer result interpretability by detecting the misleading words, as mentioned above. With the assistance of the attention mechanism, word entities in the knowledge graphs assigned with higher weights are considered candidates for misleading words. FinerFact~\cite{jin2022towards} has a different definition of interpretability. It finds claims in news content that convey incorrect information, together with user comments associated with the news post that may serve as evidence of fake or real news.

\textbf{FinerFact}~\cite{jin2022towards} constructs a \emph{claim-evidence graph}, where each node contains a claim from the  textual content of a news article together with the most relevant evidence from user comments associated with the news article that might reveal the veracity of that claim. The claim-evidence graph is fully connected to indicate all possible connections between all claims. To extract the most relevant evidence, an additional \emph{evidence graph} is constructed. The evidence graph contains the users who made the original post and comments and keywords. The connectivity of the evidence graph shows the interactions of different nodes in the graph, e.g., a user posting a comment, or comment replying to another comment,  or a term/keyword  contained in a comment, etc. Each node in the evidence graph has a saliency  score as its attribute, which indicates how important that node is in the graph. The saliency score is initialised based on meta-data, such as the number of followers that a user has, the number of times that a comment is retweeted, etc. The saliency scores on the evidence graph are propagating iteratively until they converge. 

To construct a claim-evidence graph, the topics of the news content is extracted by the LDA algorithm\cite{blei2003latent}, where each topic contains several keywords. The topics are connected to the keywords from the evidence graph to calculate saliency scores of the topics. The top-$K$ topics with the highest saliency scores are extracted, together with relevant claims and responses (found through the keywords). 
The  claims and evidence of the top-$K$ topics are then used to form $K$ nodes for the claim-evidence graph. Each node  of the this graph is classified to be fake or real, while an overall consideration of all claims is regarded as the final veracity  of the news.

\subsection{External Knowledge-Based Methods}

External knowledge-based methods draw on external knowledge to assist fake news detection. In many ways, this approach mimics the way how humans identify rumours, i.e., they seek help from authoritative sources to identify conflicts and disambiguate between fake and real news.

To connect news content with external knowledge, entity linking~\cite{shen2014entity} serves as a crucial step in nearly all external knowledge-based methods. Entities from a news article of interest are extracted using Named Entity Recognition (NER) algorithms~\cite{mikheev1999ner}. The extracted entities are then linked to external knowledge sources, such as open knowledge base, e.g., YAGO~\cite{suchanek2007yago},   Probase~\cite{song2011probase}, or Wikipedia. These knowledge bases contain a large number of 
entities (e.g., people, places, concepts, events, etc.)  with rich information as well as their   connections. Entity linking helps fetch such rich information for the entities in the news content. 
Subsequently, the external knowledge is incorporated into the knowledge graph constructed from the news content, as shown in Fig.~\ref{fig_abstraction_knowledge_driven}.

\subsubsection{Information Fusion}
Early methods such as~\cite{wang2020KMGCN,qian2021KMAGCN,li2021meet} build a knowledge graph for a news article with the help of an external knowledge base. They identify fake news by encoding the knowledge graph and fusing information from both internal knowledge and external knowledge in an end-to-end way: (1) the internal knowledge entities that are extracted are linked to external entities through entity linking; (2) a knowledge graph composed of both internal and external entities is constructed; (3) as with internal knowledge-based methods, the knowledge graph is encoded by a graph encoder, and (4) the graph representation is used for classification with a fully-connected neural network. The loss function shown in Equation~\ref{Eq.cross-entropy} is used for model training. We present three methods that adopt this approach.

\textbf{KMGCN}~\cite{wang2020KMGCN} performs fake news detection in three steps: knowledge distillation, graph construction and graph encoding. 

First, the knowledge distillation step extracts knowledge entities from the news content, including both the news text and the news images. KMGCN uses an entity linking algorithm~\cite{chen2018short} to link entities in news text to those in external knowledge bases such as YAGO~\cite{suchanek2007yago} and Probase~\cite{song2011probase}. It further uses a pre-trained YOLO-3 detector~\cite{redmon2018yolov3}
to detect objects from news images. The detected objects (i.e., their textual representations) are further linked with external knowledge bases like above. 

Next, the graph construction step creates an undirected knowledge graph with words from the news textual content and detected objects in the news images. In this graph, the edge weights are computed based on point-wise mutual information (PMI) between words corresponding to two nodes. Two nodes are connected if their PMI score is greater than 0. 

Lastly, the graph encoding step uses a GCN to learn the  representation of the constructed knowledge graph. A max-pooled node representation is fed to a fully-connected neural network to generate a news veracity prediction. 

\textbf{KMAGCN}~\cite{qian2021KMAGCN} resembles KMGCN in that it also extracts knowledge entities and models them with a graph. The difference is that KMAGCN only extracts entities from the textual content of a news article at the start. Following a procedure similar to that of KMGCN, a knowledge graph composed of textual words of the news, knowledge entities and external knowledge entities are constructed and encoded by a GCN. 
Visual features of a news article are then extracted from images through a pre-trained VGG-19 network~\cite{simonyan2014very-vgg19}. To emphasise the visual features that have better correlations with the textual information, the visual features are aligned with the knowledge graph representation using feature-level attention. In the implementation, the feature-level attention computes weights for each visual feature by comparing the visual feature with  the knowledge graph representation and re-weighting each visual feature with the computed weights. The pooled graph representation and re-weighted visual representation are then concatenated for the final classification through a fully-connected neural network. 

\textbf{LOSIRD}~\cite{li2021meet} uses  Wikipedia as the external knowledge source to verify the veracity of news articles. It has an evidence retrieval module (ERM) that is pre-trained to retrieve Wikipedia articles related to news articles of interest. The top-$K$ most relevant sentences from Wikipedia are regarded as evidence for the news. Then, a star-shaped knowledge graph consisting of the news article of interest and the evidence sentences is constructed. The news article forms the centre node, while the retrieved sentences are nodes connected to the centre node. A tree-shaped graph model reflecting the reply relationships in the news propagation process is also constructed.  The knowledge graph and the propagation graph are both encoded using the GraphSAGE model~\cite{hamilton2017inductiveGraphSAGE}, and the two graph embeddings are concatenated and used in fake news classification through a fully-connected neural network. 

\textbf{DDGCN}~\cite{Sun_Zhang_Zheng_Ma_2022} models the temporal evolution of the knowledge graph by constructing the graph gradually as more users interact with a given news article, hence more entities are extracted through user comments. DDGCN takes snapshots of the knowledge graph at different time points. The snapshots are then combined with a news propagation graph to infer the veracity of news articles, as detailed in Section~\ref{Sec.StaticGraph-based}. 

\subsubsection{Inconsistency Detection}
Some studies~\cite{hu-etal-2021-compare,fung2021infosurgeon}  predict news veracity by detecting inconsistency between the news content and external knowledge. We detail CompareNet~\cite{hu-etal-2021-compare} as a representative example of studies following this approach.

\textbf{CompareNet}~\cite{hu-etal-2021-compare} extracts knowledge entities from the textual content of news and retrieves descriptions of the entities from Wikipedia. An internal knowledge graph containing the knowledge entities, news sentences and news topics is constructed to model the news internal information. Meanwhile, external knowledge embeddings are obtained by encoding the Wikipedia entity descriptions. By comparing the internal knowledge graph representation and external Wikipedia representation, inconsistency between the news content and existing external knowledge is captured.

To construct the internal knowledge graph, a news article is first dissected into sentences. Subsequently, knowledge entities are extracted from each sentence using TAGME\footnote{https://sobigdata.d4science.org/group/tagme/}. To incorporate topic information and determine the topics conveyed in each sentence, the top-$K$ topics are identified using an unsupervised LDA algorithm~\cite{blei2003latent}. The internal knowledge graph is then composed of nodes representing sentences, knowledge entities and topics. To establish connections between these nodes, sentences are bidirectionally and fully connected to each other, while each sentence node is also linked to the entities they encompass and the topics they convey. Since the graph is heterogeneous (i.e. composed of different types of nodes and edges), a heterogeneous graph convolutional network (Hetero-GCN) is used to model the graph. In Hetero-GCN, the node representation is updated with a node-type-aware procedure as formulated by  Equation~\ref{Equ.hetero-gcn}:
\begin{equation}
 \mathbf{H}^{(l+1)} = \sigma (\sum_{\tau  \in \mathcal{T}} \mathbf{\mathcal{B}}_{\mathcal{\tau }} \mathbf{H}_{\tau}^{(l)} \mathbf{W}_{\tau}^{(l)})
\label{Equ.hetero-gcn}
\end{equation} 
where $\sigma{(.)}$ is an  activation function; $\mathcal{T} = \{\tau_{1}, \tau_{2}, \tau_{3}\}$ represents three node types: sentences, topics and entities; $H_{\tau}^{(l)}$ is the input node feature matrix of type $\tau$ at layer $l$; $W_{\tau}^{(l)}$ is the weight matrix of type $\tau$ at layer $l$; and 
$\mathcal{B}_{\mathcal{\tau}}$ is  an attention weight matrix -- its rows represent the nodes and its columns represent the neighbouring nodes of type $\tau$. After graph encoding, the  sentence and entity node embeddings are obtained. 

To retrieve external entity representations, CompareNet searches the Wikipedia page of each entity from the news content and encodes the first paragraph of the retrieved Wikipedia text as the textual representation of that entity. To capture the structural connections between all entities from a news article, TransE~\cite{bordes2013translating} is utilised to compute the entity embeddings. Then, each entity's embedding is fused from both textual and structural aspects through a gating function. 

After getting both internal and external entity representations (i.e., embeddings), CompareNet compares them to capture inconsistency, with the assumption that entities from trusted news should be better aligned with the corresponding external information. The comparison is done by calculating a comparison vector 
$\mathcal{A}_{i}$ for each entity:
\begin{equation}
    \mathcal{A}_{i} = f_{cmp}(e_{c}, W_{e}, e_{KB})
    \label{Equ.comparenet_cmp1}
\end{equation}
\begin{equation}
    f_{cmp}(x,y) = W_{a}[x-y, x\odot y] 
    \label{Equ.comparenet_cmp2}
\end{equation}
Here, $f_{cmp}(.)$ is a comparison  function, $e_{c}$ is the $i$-th entity's internal representation, $e_{KB}$ is the $i$-th entity's external representation, while $W_{e}$ and $W_{a}$ are weight matrices. 

All comparison vectors $\mathcal{A}_{i}$ are pooled using max-pooling to obtain the overall  comparison output, which is further concatenated with the max-pooled sentence outputs of Hetero-GCN before being fed into a fully-connected neural network to generate the final news veracity prediction. 

\textbf{InfoSurgeon} \cite{fung2021infosurgeon} leverages the inconsistency between textual content and the images of a news article and external knowledge. A higher inconsistency is considered a stronger hint that the news is fake. The knowledge graph construction for InfoSurgeon is similar to those in KMGCN~\cite{wang2020KMGCN} and KMAGCN~\cite{qian2021KMAGCN}, where  both internal and external knowledge are integrated in one knowledge graph. InfoSurgeon detects the nodes and edges in the knowledge graph with inconsistency. Since there is no existing dataset with such labels,  InfoSurgeon generates  
synthetic fake news samples from real news by modifying news details to inject misinformation into the knowledge graph (e.g., swapping entities, adding non-existing relationships, or replacing a sub-graph). The synthetic data is used to train a fake news detector to detect information inconsistency and hence predict the news veracity. 

\section{Propagation-based Methods}
\label{Sec.propagation-based-methods}

\begin{figure}[h]
\centering
\includegraphics[width=0.49\textwidth]{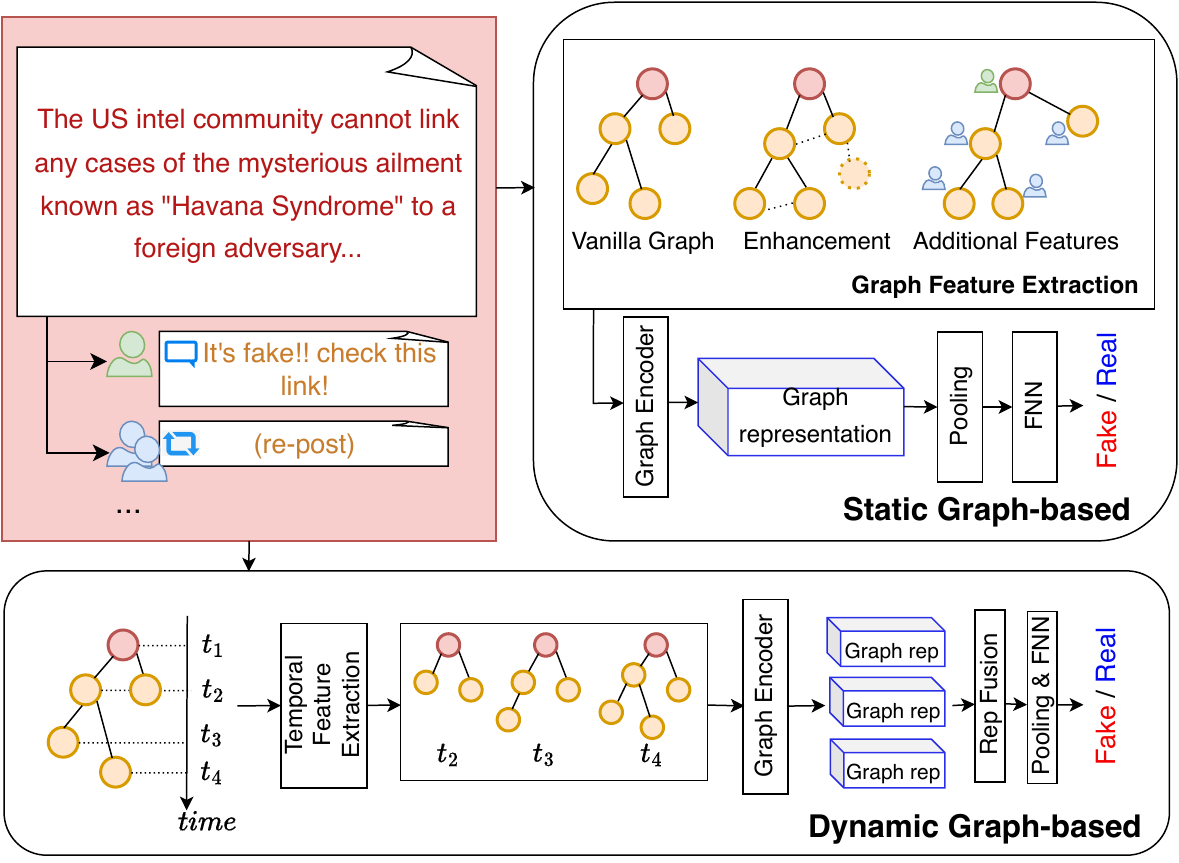}
\caption{Basic idea of propagation-based methods.}
\label{fig_abstraction_propagation-based}
\end{figure}

News in social media is directly exposed to the public, and its spread involves many social media users who interact with the news, forming unique propagation patterns. Studies  have found that fake news often spreads differently from real news, e.g., news published by official sources vs. rumours disseminated through social media~\cite{ma2016detecting,nie2020modelling}. Researchers thus have  proposed propagation-based methods aiming to exploit such differences to identify fake news. 
The propagation process involves multiple users and their interactions, e.g., comments and re-posts. Each interaction links two entities such as users, source posts and comment posts, or comments on earlier comments. These interactions form a graph structure which can together be modelled and analysed with graph-based approaches as shown in Fig.~\ref{fig_abstraction_propagation-based}.

A side benefit of considering the propagation process is that conversations formed by a news article post and its subsequent comments possess the capability to ``self-correct'' inaccurate information~\cite{ma2018rumor}. This is because users often debate and share evidence in the comments. 
Evidences hinting towards the veracity of the news may be found within the comments. This has also led to a series of studies representing the comments using tree structures and encoding them as trees/graphs for fake news detection. 

We classify and review propagation-based methods in two sub-categories: \emph{static graph-based} and \emph{dynamic-graph based}, depending  on whether temporal information is considered in the graph construction and encoding process. Beyond propagation modelling, auxiliary information such as user meta-information and advanced machine learning techniques (e.g., adversarial learning and contrastive learning) have also been utilised in propagation-based models. 

The propagation-based methods are mainly trained in a supervised way, the loss is calculated from some classification loss functions such as cross-entropy loss (Equation \ref{Eq.cross-entropy}). In addition to classification loss, some models (eg. RDEA\cite{SIGIR2021-RDEA}, GACL\cite{sun2022rumor}) utilises additional loss function to enhance the detection, which will be declared later. 

\begin{figure}[b]
\centering
\subfloat[]{\includegraphics[width=0.2\textwidth]{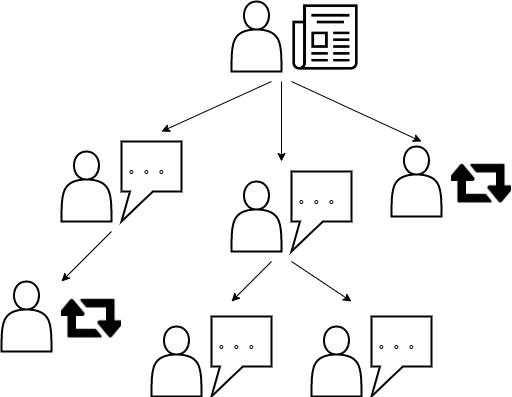}%
\label{fig_1st_prop}}
\subfloat[]{\includegraphics[width=0.2\textwidth]{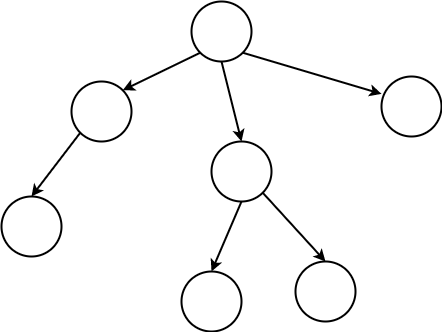}%
\label{fig_3rd_prop}}
\hfill
\subfloat[]{\includegraphics[width=0.2\textwidth]{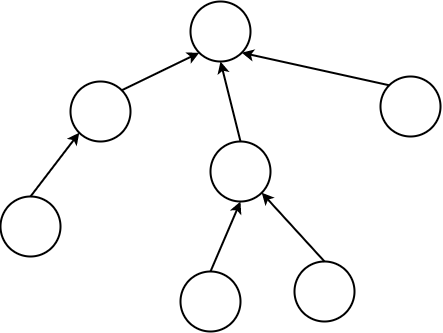}%
\label{fig_4th_prop}}
\subfloat[]{\includegraphics[width=0.2\textwidth]{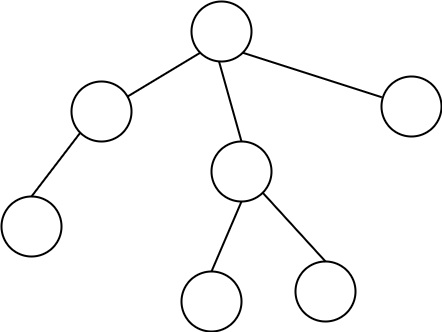}%
\label{fig_2nd_prop}}
\caption{A news propagation example and different graph modelling of the propagation process: (a) News propagation example,  (b) Top-down graph modelling, (c) Bottom-up graph modelling and (d) Undirected graph modelling.}
\label{fig_news-prop-example}
\end{figure}

\subsection{Static Graph-Based Methods}
\label{Sec.StaticGraph-based}

To capture the information flow embedded in the text of a source news post and its comments, researchers have constructed graphs using the source posts and comments as nodes, and represent the conversation through graph structures. The constructed graphs, known as \emph{propagation graphs}, are modelled using various techniques to extract the latent information, which subsequently aids in fake news detection. 
For each news item, including the user interactions (e.g., comments and reposts), a single propagation graph is constructed in static graph-based methods. The graph construction is demonstrated in Fig.~\ref{fig_news-prop-example}. For a news article and user interactions shown in Fig.~\ref{fig_1st_prop}, each interaction including the news article is represented by a node, and the nodes are connected by edges following propagation events. Depending on the edge directions, the propagation graphs can be classified into top-down directed graphs (Fig.~\ref{fig_2nd_prop}),  bottom-up directed graphs (Fig.~\ref{fig_3rd_prop}) or undirected graphs (Fig.~\ref{fig_4th_prop}). The directions of propagation are differentiated in some papers \cite{ma2018rumor,bian2020rumor,tian2022duck} to produce finer-grained propagation modelling, but the undirected graphs are the main stream in static graph-based methods. 



Next, we detail studies on static graph-based methods, starting from vanilla propagation pattern  modelling to more robust propagation modelling and then to additional feature modelling.

\subsubsection{Vanilla Propagation Pattern Modelling}

\textbf{RvNN}~\cite{ma2018rumor} models news propagation using a tree structure as illustrated in Fig.~\ref{fig_1st_prop}. A source news post and its associated comments are represented as nodes in the tree, with the source post being the root. The feature vector of each node is initialised as a textual representation of the corresponding post, i.e.,  a vector (\texttt{tf-idf} values)  of the words. A GRU-based recursive neural network (RvNN) then encodes the tree by processing nodes in both top-down (Fig.~\ref{fig_3rd_prop}) and bottom-up (Fig.~\ref{fig_4th_prop}) directions. The veracity of the source news post is classified by applying fully connected neural layers to the max-pooled tree embeddings computed by both RvNNs. It is important to note that the RvNNs for the top-down and bottom-up encoding are two separate models, referred to as TD-RvNN and BU-RvNN, respectively, and the TD-RvNN has overall better performance in the experiments. . In follow-up work, the authors of RvNN enhanced their model by replacing the GRU-based network with a Transformer encoder, resulting in the \textbf{Tree Transformer}~\cite{ma2020debunking} model with improved accuracy.

Other studies have used Graph Neural Networks (GNN) for encoding the propagation graph. For example, \textbf{TRM-CPM}~\cite{li2020exploiting} encodes the propagation graph using GraphSAGE~\cite{hamilton2017inductive}. Similar to RvNN, \textbf{BiGCN}~\cite{bian2020rumor} also models news propagation patterns with two graphs in top-down and bottom-up directions. It employs a GCN  for the top-down  graph (Fig.\ref{fig_3rd_prop}) to learn the patterns of rumour propagation, and another GCN for the  bottom-up graph (Fig.\ref{fig_4th_prop}) to capture the structures associated with rumour dispersal. The embeddings of both graphs are subsequently concatenated to classify the veracity of a news.

BiGCN has been extended in many follow-up studies with advanced GNN models. For example, Zhang et al.~\cite{zhang2021social} use  \emph{Graph Attention Networks} (GAT)~\cite{velivckovic2017graph} to model the propagation graph; others such as \cite{lin2020a,silva2021propagation2vec} use graph auto-encoders and leverage their strong learning capabilities.  GACL~\cite{sun2022rumor}  uses an adversarial contrastive learning method (discussed in Section~\ref{sec.enhanced_propagation_pattern_modelling}), which also follows the bi-directional graph modelling framework.

\subsubsection{Enhanced Propagation Pattern Modelling}
\label{sec.enhanced_propagation_pattern_modelling}

Ma et al.~\cite{ma2019detectGAN} argue that propagation graphs harvested directly may be noisy and untrustworthy. Fake news spreaders may be able to influence the graph propagation structure by deleting comments associated with their posts and/or by employing promotional bots. Additionally, news crawlers may fail to capture the entire propagation process due to privacy policies and other platform restrictions. To address these issues, Ma et al.~\cite{ma2019detectGAN} propose a \emph{Generative Adversarial Networks} (GAN)-based method to capture low-frequency but  non-trivial patterns to improve fake news detection robustness. In this model, a generator produces uncertain or conflicting comments in the propagation graph to force the discriminator to learn stronger indicative representations of the propagation patterns.

Further issues arise when news from different domains are considered, which may have  different patterns, e.g., a propagation-based model may suffer in its generalisation capability.
To address such issues, \textbf{EBGCN}~\cite{wei-etal-2021-towards} performs edge enhancement on the propagation graph that adaptively infers edge weights from node features based on the Bayes' theorem instead of treating all edges equally. It then uses GCN layers to encode the enhanced propagation graph, and the resultant graph representation is fed into a classifier for news veracity classification. Similarly, \textbf{UPSR}~\cite{wei2022UPSR} enhances graph edges using Gaussian distributions and encodes both the original and the enhanced graphs for fake news detection.

\textbf{RDEA}~\cite{SIGIR2021-RDEA} and \textbf{GACL}~\cite{sun2022rumor} follow Ma et al.~\cite{ma2019detectGAN} and  further introduce \emph{contrastive learning} into graph-based fake news detection. They train  more robust propagation-based models with contrastive learning.
RDEA presents a data augmentation strategy that randomly modifies the propagation graph (e.g.,  dropping edges or masking subgraphs). It then trains a GCN model with contrastive learning. In  contrastive learning, positive and negative data samples are fed into an encoder model in a contrastive way to encourage the model to learn data representations that better distinguish dissimilar data samples~\cite{chen2020simple-cl}. In RDEA, the modified propagations and the original propagation are regarded as a set of positive data samples, while negative data samples are obtained by random sampling from the propagation graphs with a different news class label (i.e., fake or real). 
GACL also utilises contrastive learning, and it introduce adversarial learning with an  Adversarial Feature Transformation (AFT) module. The AFT aims to improve the model's robustness to human camouflaged propagation samples. In those samples, fake news producers may manipulate the propagation to make it closer to the real instances. the AFT simulates the malicious manipulations in the model training process and forces the model to learn event-invariant features.


Another study \textbf{RDCL}~\cite{ma2022towards} finds that a minor punctuation change in the propagation graph may cause a prediction flip for  the existing models. To obtain a more robust model named, it then employs contrastive learning by  adding perturbations. The perturbations come from various aspects, including adding noise to the text representations, tuning the propagation structures as mentionedi n RDEA\cite{SIGIR2021-RDEA}.  \textbf{RDCL}~\cite{ma2022towards}.
\textbf{CCFD}~\cite{ma-CIKM2022-curriculum} further introduces \emph{curriculum contrastive learning} into fake news detection. The core idea is to gradually increase the contrastive difficulty of negative samples (i.e., making them less dissimilar to the propagation graph of interest?) to fully exploit the power of contrastive learning.

\subsubsection{Additional Feature Modelling}
Further studies attempt to extract more information from the propagation graph beyond the graph structure, e.g., characteristics of the users involved in the propagation, historical engagements of those users, and other propagation features. 


\textbf{GCAN}~\cite{lu-li-2020-gcan} collects some statistical features (eg. numbers of followers, number of posts) of all users involved in a news propagation process and constructs a fully-connected graph of the users, assuming every user has implicit connections to all others. The user features are then encoded within the fully-connected graph using a CNN encoder to serve as features for fake news detection. A limitation of this work is that the propagation graph structure is ignored and the propagation is only modelled in sequence with RNNs. 


\textbf{UPFD}~\cite{dou2021user} focuses on retweet interactions and user attributes. It constructs a graph consisting of the source news article and users who retweet the article. The user node features are extracted from their historical posts. The final graph model thus captures information related to user credibility from user historical activities through joint content and graph modeling.

\textbf{PSIN}~\cite{min2022divide} considers the \emph{propagation heterogeneity}, i.e., the different types of entities and relationships that may exist in the propagation context of a news item. It proposes a heterogeneous graph to model the propagation process, including user, post nodes and user-follower, post-reply edges. 
PSIN breaks the heterogeneous graph into three sub-graphs: a post propagation tree following BiGCN~\cite{bian2020rumor}, a user social graph formed by user-follower relationships, and a user-post interaction graph showing the authors of the posts. These three graphs are encoded synchronously by three individual neural networks in parallel, and their embeddings are concatenated to help classify the veracity of the news.

\textbf{DUCK}~\cite{tian2022duck} considers propagation information from three aspects: the structural patterns of retweets, the structural patterns of comments, and the temporal patterns of comments. Two tree-shaped propagation graphs for comments and retweets are constructed, as shown in Fig.~\ref{fig_3rd_prop}, which are encoded by two GATs. The temporal patterns of comments are described in a sequence: comments are listed in chronological order, and the list (i.e., texts) is encoded by a pre-trained Transformer encoder. The output of the GATs and the Transformer encoder are concatenated for fake news classification.

The works above primarily focus on individual propagation graphs while overlooking potential connections between multiple news sources. \textbf{UniPF}~\cite{wei-coling2022-UniPF} clusters news articles based on the K-means algorithm. News articles in the same cluster are considered to be on the same topic, and each source news node from the same topic is linked to the same topic node by an edge. As a result, the propagation graphs of different news items on the same topic are connected together through a shared topic node. Information can then be exchanged across different propagation graphs. UniPF demonstrates the potential of a hybrid fake news detection method where the connections between multiple news articles are considered, just like in the heterogeneous social context-based methods, which will be detailed in Section~\ref{sec.heterogeneous-context-based-methods}.

\subsection{Dynamic Graph-Based Methods}
\label{Sec.dynamic-graph-based-methods}

\begin{figure}
\centering
\subfloat[]{\includegraphics[width=0.25\textwidth]{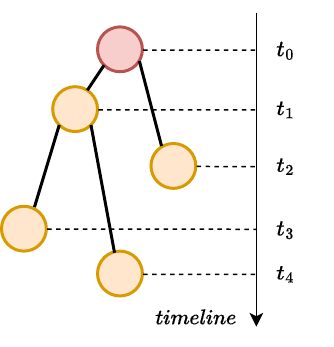}
\label{fig:a}}
\subfloat[]{\includegraphics[width=0.25\textwidth]{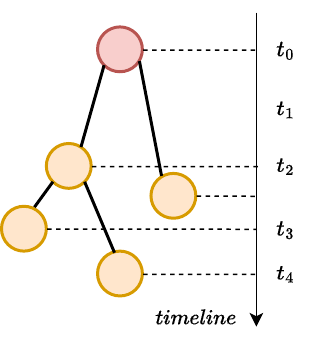}
\label{fig:c}}
\caption{Two dynamic propagation processes (a and b) with the same static pattern: From a temporal dynamic perspective, their patterns can be quite different since (a) propagates almost linearly in time while (b) has a peak at around time $t_2\sim t_3$.}
\label{fig-visualisation-dynamic-pattern}
\end{figure}

The techniques discussed in the previous subsection assume a complete (static) news propagation graph. In reality, the full propagation structure of news does not emerge instantaneously. Instead, it undergoes an evolution process, expanding from one or two nodes to potentially a much larger propagation graph. Such a temporal process has been omitted (or significantly simplified) by static graph-based methods. This misses important patterns for fake news detection. 
For example, in Fig.~\ref{fig-visualisation-dynamic-pattern}, two news propagation graphs (where each node represents a post) have identical tree structures, but their propagation patterns  differ when temporal information is  considered.

It has been observed that fake news propagation exhibits a viral pattern with multiple stages reflecting  people's attention and reactions. This gives rise to a distinct life cycle~\cite{shu2020fakenewsnet}. Studies exploit such patterns for fake news detection with dynamic propagation graph-based methods.

\textbf{DSTS}~\cite{ma2015detect} emphasises the importance of temporal features and proposes a feature engineering-based method that considers features such as the number of retweets over time when detecting fake news. 
\textbf{MMHM}~\cite{naumzik2022detecting} models the retweeting dynamics using marked Hawkes processes~\cite{hawkes1971spectra} based on the \emph{self-exciting phenomenon}~\cite{hawkes1971spectra}. The self-exciting phenomenon refers to the occurrence of an event that can influence the likelihood of future events. In the social media context, this can involve a phenomenon whereby some user interactions (e.g., comments, reposts) can cause a sudden increase or decrease in the propagation of a news item. MMHM demonstrates the potential of fake news detection using only self-exciting phenomenon features and opens up new research opportunities.

\textbf{RDLNP}~\cite{lao2021rumor}  learns the temporal sequential characteristics of the propagation patterns, by adding linear sequence learning alongside graph-based structural learning. To simplify the model, only an undirected propagation graph was considered (cf.~Fig.~\ref{fig_2nd_prop}). As before, a GCN is used to encode the graph structure, while an LSTM network is used to encode the linear sequence. The graph and sequence embeddings are then integrated using an attention mechanism for fake news classification. 

\textbf{Dynamic-GCN}~\cite{choi2021dynamic} and \textbf{Dynamic-GNN}~\cite{song2022dynamic} capture temporal dynamics by taking snapshots of the propagation graph at different time points. These graph snapshots are encoded by a GNN (e.g., a GCN) to generate graph representations, which are then processed using sequence modeling approaches, such as self-attention encoders. Similarly, \textbf{DDGCN}~\cite{Sun_Zhang_Zheng_Ma_2022} extracts temporal graph snapshots. In addition, it extracts knowledge entities and builds temporal snapshots of the resultant knowledge graphs. The temporal snapshots of both the propagation graphs and the knowledge graphs are encoded simultaneously and interactively for fake news detection.

\textbf{TGNF}~\cite{song2021temporally} considers temporal information at the graph node level. It utilises a temporal pattern-aware GNN named \emph{Temporal Graph Network} (TGN)~\cite{rossi2020temporal} to encode the evolution of propagation graphs. In the node aggregation step of TGN, a node's active time and historical status are considered. The TGN updates node features according to the temporal aspects of node interactions. Given an interactionbetween two nodes, only the two node embeddings are updated using the attention mechanism.  TGNF utilises TGN to build a graph structure based on the sequence of user interactions (i.e., replies and retweets on social media). The  graph is encoded by a GCN, which is then fed into a fully-connected neural network to predict the news veracity. 

\textbf{UGRN}~\cite{chen2022progressive} introduces the concept of \emph{trigger detection} to identify user interactions (retweets or comments) that yield an marked increase in propagation pattern of the source news item. If the triggers are found properly, they can be used to infer the  trustworthiness of the source news based on the trigger information. After applying GCN to the propagation graph, the node representations are listed in chronological order. A classifier is then trained at the node level to classify whether a node is a trigger. Meanwhile, UGRN also feeds the graph representation to a fully-connected neural network to predict whether the news is real or fake. The trigger detection and fake news detection models are jointly trained as a multi-task learning task using the following loss function. 
\begin{equation}
    \mathcal{L} = \mathcal{L}_{t} + \mathcal{L}_{v}
\label{Equ.trigger_formula}
\end{equation}
where $\mathcal{L}_t$ and $\mathcal{L}_{v}$ are the cross-entropy loss for trigger classification and news trustworthiness classification, respectively. Both of $\mathcal{L}_t$ and $\mathcal{L}_{v}$ are calcualted based on Equation.\ref{Eq.cross-entropy}. 

Similarly to UGRN, \textbf{SEAGEN}~\cite{gong2023fake} also encodes the propagation process with a graph encoder, and then it predicts interaction intensities over the propagation. The intensity here is the likelihood for the source news to receive an interaction at a given time, which reflects the self-exciting phenomenon. SEAGEN is also trained in a joint manner, in which the cross-entropy loss for news veracity and intensity accuracy loss are combined together as the overall model loss. The intensity accuracy  loss is calculated by comparing the real propagation intensity and the estimated one. Both UGRN and SEAGEN consider the graph propagation speed. The difference between them is that UGRN aims to find the trigger that invokes a sudden increase in the retweets or comments, while SEAGEN aims to model the evolving propagation speed using the self-exciting phenomenon.

\section{heterogeneous social context-based Methods}
\label{sec.heterogeneous-context-based-methods}

Beyond the textual content of news posts and their associated comments, other entities such as the users making those posts and comments, and the topics expressed by the posts can be considered as the basis of the  veracity of news articles. Such additional information form a heterogeneous social context that reveals implicit connections among multiple news. For example, multiple news may be posted or spread by the same users. This section focuses on graph-based methods that model such heterogeneous social contexts through \emph{heterogeneous graphs}. A heterogeneous graph is a graph that consists of  more than one type of nodes or  edges. After a heterogeneous graph is  constructed, graph embedding techniques can be used to encode the graphs for fake news classification. 

To extract information from heterogeneous graphs, existing methods either extract a sub-graph most relevant to a news post and encode the sub-graph with a graph encoder such as GCN \cite{kipf2017semisupervised-gcn}, or they construct a large graph with representation optimisation and then take the optimised representation for veracity classification. Depending on whether the sub-graph-level features or the node-level features are utilized in the detection process, we classify the heterogeneous social context-based methods into two sub-categories: \emph{sub-graph-level classification} and \emph{node-level classification}. Intuitively, node-level methods run graph algorithms on the full heterogeneous graphs, while sub-graph-level methods extract a sub-graph and only run a graph encoder on the sub-graphs. Fig.~\ref{fig_abstraction_propagation-based} illustrates both types of method. 

\begin{figure}[h]
\centering
\includegraphics[width=0.49\textwidth]{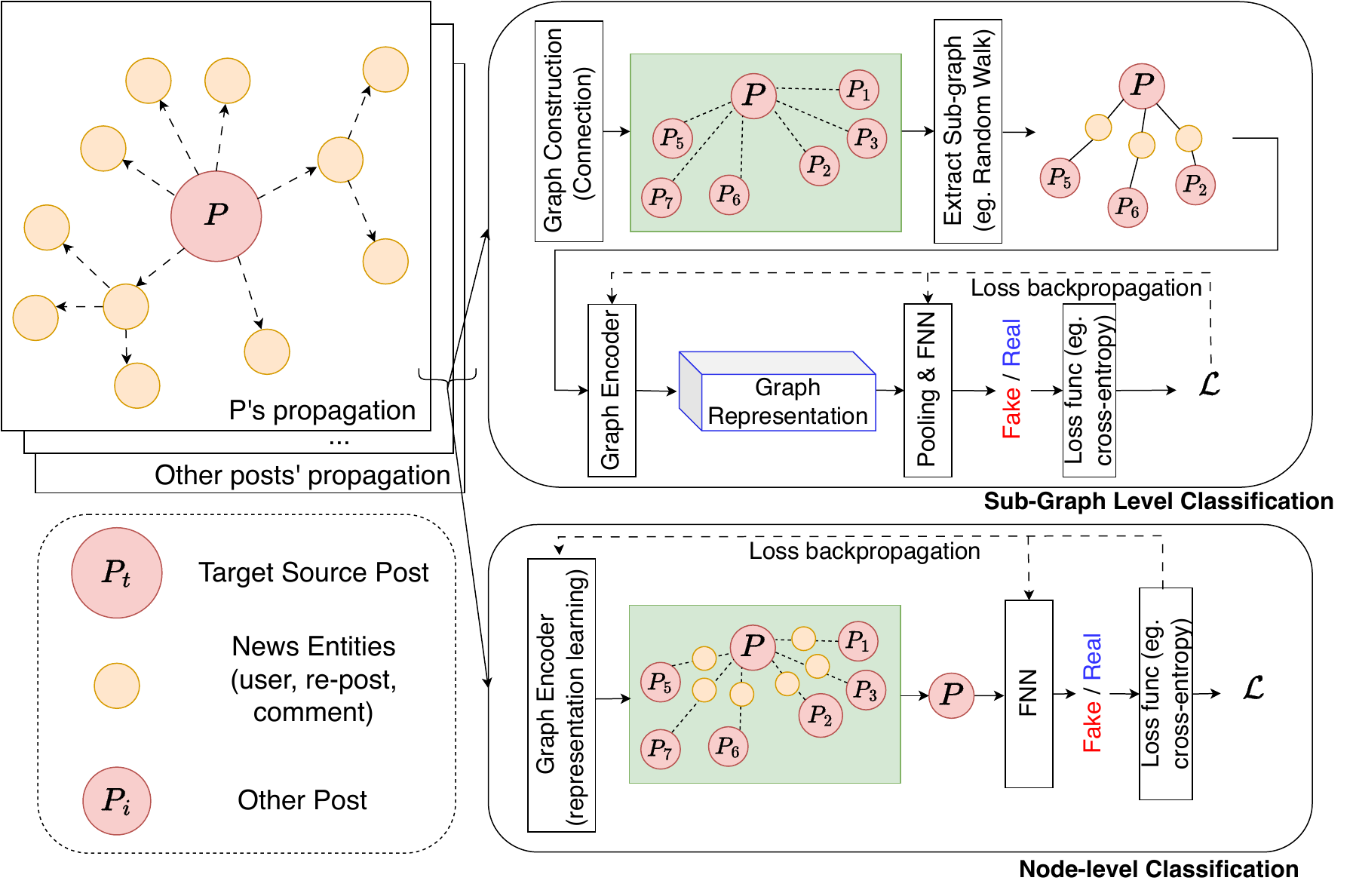}
\caption{Basic idea of heterogeneous social context-based methods. }
\label{fig_abstraction_propagation-based}
\end{figure}
 
\subsection{Sub-graph-level Classification}

Sub-graph-level methods construct a large heterogeneous graph composed of all news items, users and user interactions (e.g., comments or user-follower relationships). Then, a sub-graph is extracted from the large graph for each news item by sampling the k-hops neighbours of it, to be encoded by a heterogeneous graph encoder. The encoded sub-graph embedding is fed into a fully-connected neural network for news veracity classification. This process is similar to the propagation-based methods described in the last section. A core difference is that propagation-based methods construct a propagation graph for each news item, neglecting the implicit connections between the news item and other news items. In contrast, sub-graph-level methods reviewed in this section consider a wider social context where news items are  connected to each other, through either user interactions or topic similarity. 

Two early methods,  \textbf{Monti}~\cite{monti2019fake} and \textbf{GLAN}~\cite{yuan2019GLAN},  construct a large heterogeneous graph over all news articles and users. They model the user-follower relationships and news propagation relationships as edges in the graph. To detect fake news, Monti extracts sub-graphs corresponding to news article of interest and encodes the sub-graph by a GCN. The mean pooling of the node representations computed by the GCN is then fed into a classifier for news veracity classification. GLAN follows a similar procedure but uses a CNN and multi-head attention as the graph encoder.

\textbf{NDG}~\cite{kang2021fake} uses the nodes of the heterogeneous graph to represent news comments and news domains. 

The edges in the graph represent users commenting on a news article and news articles belonging to a given domain. There may be too many edges connected to a node in such a graph (e.g., too many comments on the same news article or too many news articles on the same domain=Î), resulting in a ``neighbourhood explosion'' problem. To mitigate this problem, a sampling strategy that randomly takes a subset of the graph is proposed in NDG to retain a size-limited sub-graph. The rest of the classification process is the same as in Monti and GLAN.

\textbf{MFAN}~\cite{mfan2022zheng} captures unobserved links in the propagation graph. It models the  social propagation context with graphs that contain three types of nodes: the news article node,  user comment nodes, and user nodes (i.e., users who made comments). MFAN argues that the links and relationships in the graph structure may be incomplete due to privacy issues or due to manipulation by fake news spreaders. 
MFAN then complements the graph links by inferring hidden links. During the inference, MFAN compares the cosine similarity between every pair of nodes in the graph and assigns edges to the pairs with high similarity. Then, an adaptive GAT is utilised to encode the graph to obtain the graph embedding. 

\textbf{Gregor}~\cite{donabauer2023exploring} constructs a heterogeneous graph to model news propagation processes. The news article and the users who posted about the article  are linked together in a graph that is encoded by a GNN. 
Similarly, \textbf{SureFact}~\cite{yang-kdd2022-reinforcement} constructs a heterogeneous propagation graph consisting of four types of nodes corresponding to the news article, users, user posts (retweets and replies), and keywords (extracted through topic models such as the LDA algorithm~\cite{blei2003latent}). Then, SureFact uses reinforcement learning to select the most important sub-graphs for fake news detection, since considerable noise and irrelevant information may exist in the full graph. 

After filtering the irrelevant information, the remaining sub-graphs are modelled by an adapted GAT in a fine-grained manner, to improve the model's discrimination power and explainability. Compared to the earlier graph-based methods that capture mainly high-level patterns of the full propagation graph, SureFact provides deeper insights that can be used to understand the propagation patterns of fake news through filtered sub-graphs. 

Chandra et al.~\cite{chandra2020graph} build a large heterogeneous graph consisting of users and news articles. They detect fake news via analysing the user communities (i.e., user clusters). Their core idea is that a news's veracity would be similar to those  topologically close to that news. The veracity of a news article thus may be inferred from other news articles in the same cluster that come with a veracity label. 

\subsection{Node-level Classification}

Other studies construct a large heterogeneous graph and exploit labelled news article nodes to help classify unlabelled ones. Such methods are referred to as node-level methods. In such methods, all news and users participate in the propagation process are considered as nodes in the graph, and the connections among news are represented by edges. The graph representation is optimised by GNNs during graph construction. The learned embedding of each news node can be  utilised for news veracity prediction directly, i.e., no sub-graph extraction is needed. 

\textbf{TriFN}~\cite{shu2019beyond}  is the first approach to model multiple news articles in a large heterogeneous graph. 
In TriFN, news articles are connected to their publishing users as well as spreading users. Such a structure has been adapted in many follow-up studies~\cite{nguyen2020fang,kang2021fake,mfan2022zheng,mehta2022tackling}.
Both news articles with and without veracity labels are included in the heterogeneous graph. The graph representation is then optimised through Laplacian matrix decomposition. The unlabelled articles are then classified by a classifier trained on the labelled articles.

\textbf{FANG}~\cite{nguyen2020fang} enhances the heterogeneous graph representation by a node representation proximity loss below, which is optimised together with the news veracity classification loss based on Equation~\ref{Eq.cross-entropy}: 
{\small
\begin{equation}
    \mathcal{L}_{prox} = - \sum_{r \in G} \left [ \sum_{r_{p} \in P_{r}} \log(\sigma (z_{r}^{T} z_{r_{p}}) ) + 
Q\sum_{r_{n} \in N_{r}}  \log(\sigma (-z_{r}^{T} z_{r_{n}}))  \right ] 
\label{Eq.proximity_loss}
\end{equation}
}
Here $z_{r} \in \mathbb{R}^{d}$ is the representation of entity $r$ in the heterogeneous graph $G$, $P_{r}$ is the set of nearby nodes (\emph{positive set}) of $r$, and $N_r$ is the set of disparate nodes (the  \emph{negative set}) of $r$. Set $P_r$ is obtained with a random walk from the entity node $r$, and $N_r$ is derived using negative sampling over the graph. The proximity loss optimises the graph representation based on the \emph{echo chamber} effect, in which nodes closely connected within the same community share similar features, while those that are farther apart in distinct communities exhibit different features.
Overall, fake news detection in FANG is modelled as a reasoning problem over a graph where the graph representation is refined in parallel. FANG uses the evidence provided by existing knowledge of real and fake content from the training data to assess the authenticity of unknown news (i.e., test data) based on the observed links in the heterogeneous graph. 

Following TriFN and FANG, a series of other methods have been proposed~\cite{ren2020adversarial,yuan-etal-2020-early} that use advanced machine learning techniques such as adversarial active learning and weakly supervised learning for node-level fake news classification.  


Based on the graph representation of FANG, \textbf{Mehta}~\cite{mehta2022tackling}  enriches the edges in the graphs. Its idea is similar to MFAN~\cite{mfan2022zheng} but from a node-level classification perspective. Since there are multiple types of nodes in the heterogeneous graph, different inference operators are used to augment the graph with different relationships beyond those initially seen. Fake news detection in Mehta is performed in a transductive manner using a \emph{Relational Graph Neural Network} (R-GNN)~\cite{schlichtkrull2018modeling}. The information in the observed data is transferred to the veracity-unknown nodes, for veracity prediction.  

\textbf{Hetero-SCAN}~\cite{cui-CIKM2022-HeteroSCAN} decomposes the heterogeneous graph into two levels. The higher level describes the correlations between news publishers and news articles, while the lower level describes the correlations between news spreaders and news articles. The two-level heterogeneous graph is divided into sub-graphs named \emph{meta-paths}, to model the detailed correlations, e.g., a news article cites another, and a social media user spreads two news articles. During fake news detection, a news article's surrounding meta-paths are extracted, encoded with GRU and attention encoders, and integrated based on the attention mechanism to assess the veracity of the news article. 

\textbf{TR-HGAN}~\cite{gao2022topology} considers the  \emph{topology imbalance} and \emph{relation inauthenticity} when inferring the veracity of a news article. Here, the topology imbalance means that the topology distribution of labelled news nodes is likely to be asymmetric and uneven. To address this issue, TR-HGAN proposes a smoothing strategy to force the influence from labelled nodes to decay with the topological distance Relation inauthenticity refers to the fact that the propagation structure is not always reliable because it can be manipulated by users and thus are unauthentic. To address this issue, TR-HGAN adapts a hierarchical attention mechanism that weights a node's neighbours based on the nodes and their types (e.g. user nodes, news nodes or comment nodes) to deal with propagation graphs being manipulated by fake news spreaders.

\section{Comparison of Reported Empirical Model Performance Results}
\label{sec.dataset}
Various datasets have been utilised in the  experiments of the existing fake news detection studies. 
Below, we summarise the most commonly used datasets:
\begin{itemize}
    \item \textbf{Twitter15}~\cite{ma2017detect}  includes popular source tweets that were highly retweeted or replied (in 2015) along with the  propagation graphs (retweets and replies). 
    \item \textbf{Twitter16}~\cite{ma2017detect} shares the same data collection process withTwitter15 but was based on data from 2016.  
    \item \textbf{PHEME} has two versions: PHEME-5~\cite{zubiaga2016learning-PHEME5} and PHEME-9~\cite{Kochkina2018-PHEME9}.  PHEME-5 was collected in 2016 from five news events covering five domains, whilst PHEME-9 was collected in 2018 from nine news events (and domains).  Only replies are collected in the propagation graphs. 
    \item \textbf{Weibo}  \cite{ma2016detecting} contains news propagation graphs crawled from the Weibo platform. Similar to Twitter15 and Twitter16, source news, reposts and replies are collected. 
    \item \textbf{FakeNewsNet}~\cite{shu2020fakenewsnet} was composed of two topics: politics and entertainment. Rich social context information is included such as the news, news sources, news spreaders,  retweets and reply propagation graphs, etc. FakeNewsNet is also divided into two smaller datasets: \textbf{PolitiFact} and \textbf{GossipCop} depending on the news topics. 
\end{itemize}

Most of these datasets are obtained from Twitter. Social media platforms like Twitter and Weibo have become mainstream data sources for social media research in recent years, due to their scale, adoption and openness of data. The statistics of these datasets is shown in Table~\ref{Tab.dataset_feature}. 

\begin{table*}[h]
\centering
\caption{Statistics of Datasets}
\label{tab:my_label}
\begin{tabular}{l|c|c|c|c|c|c|c}
\toprule
\textbf{Feature} & \textbf{Twitter15} & \textbf{Twitter16} & \textbf{PHEME-5} & \textbf{PHEME-9} & \textbf{Weibo} &\textbf{Politics} &\textbf{GossipCop} \\ \midrule
Number of source news& 1,490              &  818               &   5,802          & 6,425            & 4,664  &1,056 & 22,140    \\ \hline
Number of users      &   276,663          &  173,487           &   49,435         & 50,593           & 2,746,881  & 345,440 & 345,292   \\ \hline

Number of posts      & 331,612            & 204,820            &   103,212        & 105,354          & 3,805,656   &564,129 & 1,396,548 \\ \hline
Number of classes    & 4                  & 4                  &   2              & 2                & 2    & 2 &2        \\ \hline
Number of fake news  &  -                 &  -                 &   1,972          & 3830             & 2,313   & 432 & 5,323     \\ \hline
Number of real news  & -                  & -                  &   2402           & 4023             & 2,351   & 624 & 16,817     \\ \hline
Number of non-rumors & 374 & 205          & -                  &  -               &  -               &  - & -         \\ \hline
Number of false rumors & 370  & 205       & -                  &  -               &  -               &  - &-          \\ \hline
Number of real rumors & 372 & 205         & -                  &  -               &  -               &  - &-          \\ \hline
Number of unverified rumors & 374 & 203      & -                  &  -               &  -               &  - &-          \\ \hline
Average number of time length/news & 1337 hours & 848 hours       &  -                &  -                & 2,461 hours & - & -\\ \hline 
Average number of post/news & 223            & 251                & -                 & -                 & 816         & - & -   \\ \hline
Maximum number of posts/news & 1,768          & 2,765              & -                &  -                 & 59,318      & - & -     \\ \hline
Minimum number of posts/news & 55 & 81 & -  & - & 10 & - & -\\ \bottomrule
\end{tabular}
\label{Tab.dataset_feature}
\end{table*}

\begin{table*}[h]
\centering
\caption{Accuracy  of Different Models on the Different Datasets.}
\label{tab:my_label}
\begin{tabular}{l|c|c|c|c|c|c|c}
\toprule
\textbf{Method} & \textbf{Twitter15} & \textbf{Twitter16} & \textbf{PHEME-5} & \textbf{PHEME-9} & \textbf{Weibo} &\textbf{Politics} &\textbf{GossipCop} \\ \midrule
\multicolumn{8}{c}{\textbf{Knowledge-driven Methods}}  \\ \midrule
FinerFact\cite{jin2022towards}& -& -& -& -& -& 0.920 (\#815) & 0.862 (\#7,162) \\ \hline
KMGCN\cite{wang2020KMGCN} & - & - & 0.876 (\#5802) &- & 0.886 (\#4664) & - & -\\ \hline
KMAGCN\cite{qian2021KMAGCN}  & -& - & 0.867 (\#5802)&- & 0.944 (\#9528) & - & - \\ \hline
LOSIRD\cite{li2021meet} &- &- & \textbf{0.914} (\#5802) & \textbf{0.925} (\#6425) & - & -\\ \midrule
\multicolumn{8}{c}{\textbf{Propagation-based Methods}}  \\ \midrule
RvNN\cite{ma2018rumor} & 0.723 (\#1490) & 0.737 (\#818) & - & - &- &- & -\\ \hline
TRM-CPM\cite{li2020exploiting} & -& -& 0.900 (\#5802) & 0.919 (\#6425) & - & - & -\\ \hline
Bi-GCN\cite{bian2020rumor}  & 0.886 (\#1490) & 0.880 (\#818) & - &- & 0.961 (\#4664) & - & -\\ \hline
RNLNP\cite{lao2021rumor} & - & - & - & 0.919 (\#3164) & - & - & -\\ \hline
EBGCN\cite{wei-etal-2021-towards} & 0.892 (\#1490) & 0.915 (\#818) & -& 0.715 (\#2402) & - & - & -\\ \hline
UPSR\cite{wei2022UPSR} & -& -& -& -& -& \textbf{0.914} (\#314) & \textbf{0.977} (\#5464)\\ \hline
EDEA\cite{SIGIR2021-RDEA} & 0.855 (\#1490) & 0.880 (\#818) & - & - & - & - & -  \\ \hline
GACL\cite{sun2022rumor} & 0.901 (\#1490) & 0.920 (\#818) & -& 0.850 (\#6425) & - & - & -\\ \hline
RDCL\cite{ma2022towards} & -& -& 0.871 (\#5802) & 0.864 (\#6425) & - & - & - \\ \hline
CCFD\cite{ma-CIKM2022-curriculum} & 0.856 (\#1490) & 0.886 (\#818) & - & - & 0.975 (\#4532) & - & -\\ \hline
UPFD\cite{dou2021user} & - & -& -& -& -& 0.846 (\#314) & 0.972 (\#5464)\\ \hline
DUCK\cite{tian2022duck} & 0.900 (\#1490) & 0.910 (\#818)&- & - & \textbf{0.980} (\#4664) & - & -\\ \hline
UniPF\cite{wei-coling2022-UniPF} & 0.959 (\#712) & 0.963 (\#410) &- &- &- & 0.911 (\#314) & 0.966 (\#5464) \\ \hline
Dynamic-GCN\cite{choi2021dynamic} & 0.827 (\#1490) & 0.836 (\#818) &- &- & 0.936 (\#4664) & - & -\\ \hline
Dynamic-GNN\cite{song2022dynamic} &- & - & -& -& 0.957 (\#4338) & - & -\\ \hline
TGNF\cite{song2021temporally} &- & - & -&- & 0.968 (\#4338) & - &-\\ \hline
DDGCN\cite{Sun_Zhang_Zheng_Ma_2022} &- & - & 0.855 (\#4657) &- & 0.948 (\#5748) & - & -\\ \midrule
\multicolumn{8}{c}{\textbf{Heterogeneous Social Context-based Methods}}  \\ \midrule
NDG\cite{kang2021fake} & - & - & -&- & 0.961 (\#4197) & - & -\\ \hline
SureFact\cite{yang-kdd2022-reinforcement} &- &- &- &- &- & \textbf{0.9413} (\#815) & 0.8797 (\#7612) \\ \hline
TR-HGAN\cite{gao2022topology}& \textbf{0.929} (\#1490) & \textbf{0.932} (\#818) &- &- & 0.963 (\#4664) & - & -\\ \bottomrule

\end{tabular}
\label{Tab.model_performance}
\end{table*}

The detection performance (i.e., accuracy) of the  models reviewed over the  datasets above (as reported in the original papers) is summarised in Table~\ref{Tab.model_performance}. Note that even when the same datasets are utilised, different papers may run experiments with  different subsets. Therefore, we label the size of the datasets in the performance result table with a '\#' symbol. For example, $0.876 (\#5802)$ means the model accuracy is 0.876 when experimented on a subset of 5,802 source news. If two papers utilised  subsets of the same size, we consider that they have the same experiment setting and hence are comparable even though the dataset splits for training, development and testing may still be slightly different. 

As seen in Table~\ref{Tab.model_performance}, propagation-based methods are the primary approaches utilising the above datasets. Many methods perform experiments with different versions of the datasets. For example, the Weibo~\cite{ma2016detecting} dataset has many versions in the table as the dataset size varies from 4,197 to 9,528 source posts. The different versions come from filtering and re-sampling to make the dataset more suitable for the given experiment. For example TGNF~\cite{song2021temporally} and Dynamic-GNN~\cite{song2022dynamic} exploit temporal features to detect fake news, but some samples from Weibo are too short to be used for such temporal features which are hence  filtered. In addition, FakeNewsNet (PolitiFact and GossipCop)~\cite{shu2020fakenewsnet} is released based on Twitter post tokens due to the privacy concerns. To get the full content, researchers need to crawl Twitter data using these tokens. However, since the contents on social media platforms are dynamic, it can be the case that some posts and users are not available any more at a later date. Therefore, the size of FakeNewsNet may be reduced in the later papers.

In terms of model performance comparison, we  observe that models such as LOSIRD~\cite{li2021meet}, UPSR~\cite{wei2022UPSR}, DUCK~\cite{tian2022duck}, SureFact~\cite{yang-kdd2022-reinforcement} and TR-HGAN~\cite{gao2022topology} achieve the best performance over these datasets, respectively. LOSIRD uses  knowledge from Wikipedia  for debunking fake news. UPSR questions the reliability of propagation patterns because the propagation graph can be deliberately perturbed and imbalanced, therefore it proposes a Gaussian distribution-based edge enhancement to make the propagation-based detection more robust. The strength of DUCK is that it considers the propagation pattern from multiple aspects including  linear and non-linear aspects, static and temporal aspects, as well as retweets and comments. SureFact and TR-HGAN both utilise heterogeneous graphs to model the social context of news. SureFact uses reinforcement learning to select informative parts of the social context graph to filter noise. TR-HGAN improves the robustness of heterogeneous social context-based methods by solving graph topology imbalances and relation inauthenticity issues. 

\textbf{Limitations of existing datasets and experiments.} 
Though much progress has been achieved, important issues persist:
\begin{enumerate}
\item A standard benchmark is currently lacking. Different models are implemented on different datasets and/or different versions, leading to challenges in assessing and comparing existing works. 

\item Some of the existing datasets are also outdated. With the exception of FakeNewsNet, the other datasets were released at least five years ago. Considering the constantly changing news interests and sharing patterns, it is less meaningful to design models on historic datasets. For example, nowadays AI models are able to synthesise fake news, but no dataset exists to train models to detect AI-synthesised misinformation. 

\item The detection performance related to accuracy and F1 scores should not be the only evaluation metrics. Other attributes such as early detection as well as cross-domain detection capability should also be considered. Most  existing methods are trained on the full time span of the samples. This means that rich information and evidence are provided to the models to judge if a news is fake. However, such full-time-span information is not always available. In reality, if some news propagates explosively in a short  period of time, the negative impact may already be done to the society and any individuals involved. Therefore, models should have the capability to discover fake news from different topics and  proactively intervene to stop the spread of fake news early on. 
\end{enumerate}

\section{Challenges and Opportunities}
\label{sec.challenges-and-opportunities}

As reviewed above, many  automatic graph-based fake news detection methods have been proposed. However, few of them are have been deployed in real application systems. For example, with Twitter, humans are still hired  as fact checkers to address fake news\footnote{https://help.twitter.com/en/resources/addressing-misleading-info}. The gap between the variety of automatic detection methods and real-world deployment is a major shortfall of existing studies. 

We consider several key issues from different aspects based on the reviewed graph-based fake news detection methods: \emph{explainability}, \emph{cross domain detection}, \emph{real time detection}, and \emph{efficiency and cost}. 


\subsection{Explainability}


\emph{Explainability} in fake news detection refers to the ability  to explain why a news article might be considered as misinformation or fake news. Some information in the fake news arena may be incorrect which can be shown directly by referring to authoritative knowledge sources, whilst other news may be posted and spread by propaganda social media accounts that may have elements of truth or be a small part of a  more complex news item used to influence the public, etc. 

Existing work tends to answer such explainability challenges based on model interpretability. Interpretability here refers to the ability to understand and make sense of the internal workings of a machine learning model \cite{molnar2020interpretable-book}. It focuses on the model's internal mechanisms, such as the relationships between the input features and the mapped output predictions. By providing model interpretability, humans should be able to comprehend and reason about how and why the model classified a given news item as fake or real.

Interpretability has been  defined in different ways in the literature.  The seminal work~\cite{zhou2019network} shows how high-level statistics about the propagation networks (e.g., community density) are important for fake news detection. Another work~\cite{khoo2020interpretable} defines the interpretability as the ability to identify the user comments that hold objections or contrary opinions and can reveal misinformation in the source posts. This is based on the self-correction ability of  propagation, i.e.,  fake news usually receives more objections and alternative opinions than real news.  Other studies~\cite{Xu-WWW22-GET,fu-cikm2022-disco} define interpretability of news content based on capturing misleading words. 


Many attempts to offer interpretability for fake news detection have  exploited  attention mechanisms \cite{vaswani2017attention}. The attention mechanism weights elements (e.g., users, posts, or words) in an input sequence and graph as the basis for the model decision making (i.e., classifying a news item as being fake or not).  Though much progress has been made,  the gap between  model interpretability and real explainability remains substantial. 


Some models introduce more complex interpretability approaches. SureFact~\cite{yang-kdd2022-reinforcement} considers the information of sub-graphs in a propagation graph to be  imbalanced, meaning that some sub-graphs can contain useful information like anomaly patterns that can be used to reveal the veracity of news, while other sub-graphs cannot. To select more informative sub-graphs, a reinforcement learning-based selector is designed and aligned with the fake news detection process. By showing  anomaly patterns in sub-graphs, SureFact explains why a news has been  classified as fake.  FinerFact~\cite{jin2022towards} takes a knowledge-based approach. It extracts  potential fake claims from a news article and supporting evidences from user comments. By analysing the saliency of the evidence and graph modelling, FinerFact explains which claims in the news might be fake and attempts to provide supporting evidences. 

SureFact \cite{yang-kdd2022-reinforcement} and FinerFact  \cite{jin2022towards} offer fake news detection reasoning based on the propagation graph and textual contents. Further work is required. On the one hand, existing explainability work is mainly driven by attention mechanism weights. Words in texts or nodes in propagation graphs with higher weights are assumed to be more important, however a semantic connection is lacking in this process. Even with the correct veracity prediction and accurate attention weights, it may not be persuasive enough for people to believe in the the explanation of the result. Witnessing the success of semantic reasoning of large language models such as ChatGPT, the explainability of fake news detection can also be improved from a semantic level. One the other hand, when people reason about  news veracity, the broader social context needs also to be considered. For example, the social context includes what is posted concurrently by other users, what are the connections between those users, what is the verified knowledge background, etc. Reasoning based on the broader social context is still under-explored and an area of future work. 


\subsection{Cross-domain Detection}

Fake news detection has unique challenges for model generalisability. Textual content plays a key role in deciding whether a news is fake, while the content distribution can vary substantially across different news domains. A model trained for news in one domain may suffer from a substantial performance loss (around 30\%~\cite{min2022divide}) when being applied to fake news detection in another domain. A key challenge is how to obtain a generalisable fake news detection model. 

Early studies \cite{ma2018rumor,bian2020rumor} have shown that fake news and real news exhibit different propagation patterns, and they try to avoid the entity-bias from individual domains by learning specific propagation patterns of fake news. Data augmentation strategies \cite{SIGIR2021-RDEA}  are often used to enhance the learning of propagation patterns and improve model robustness.  However, a recent study~\cite{min2022divide} analysed the propagation of news from different domains and found that the propagation pattern also suffers from  entity-bias. Therefore, the study utilises multi-task learning to detect fake news and classify news topic contemporaneously. The assumption is that if a model can identify the topic of news, it can capture topic-agnostic features to mitigate the domain dependence of the trained model. 

Another possible solution to the entity-bias problem is to train models continuously by techniques such as active learning, i.e., keeping the models updated with fresh knowledge from real-time news sources. An initial study~\cite{yue-cikm2022-contrastive} was done on this idea and demonstrated its feasibility. However, the study only experimented with NLP models, while graph-based methods remain under-explored. 

At present, cross-domain fake news detection is mainly studied under individual news settings that consider the content and context of of a news item. When considered from a broader perspective, it has been observed that different rumours propagate concurrently, even when some rumours have already been shown to be fake. When judging the veracity of a single news item, we can consider information from other concurrent news (e.g., similar information is being shared by unverified users), and historical news (e.g., some users who have been previously detected to spread fake news). Such information can be used to improve fake news detection. In summary, capturing a broader  social context also helps to understand the overall news context, and can contribute to a better understanding of news from various domains.

\subsection{Real-time Detection}

The need for real-time detection of fake news cannot be overstated. Swift identification and prevention of the dissemination of fake news is crucial. However, real-time detection remains a significant challenge for current methods due to three limitations: weakness in temporal modelling; lack of connections between multiple social media platforms and the fact that any solution has to be accepted and rolled out by the major social media platforms themselves. Note that for the last point, this may not be in the business interests of the platforms, e.g., fake news is often the source of attraction and engagement to many user communities.

It is the case however that the majority of models discussed in this survey are built upon static graphs. These include the static propagation-based models discussed in Section \ref{Sec.StaticGraph-based}, the knowledge-driven models outlined in Section \ref{sec.knowledge-driven}, and the heterogeneous social context-based models presented in Section \ref{sec.heterogeneous-context-based-methods}. It is worth mentioning that utilising static graphs results in the loss of (current) temporal information and hence may not effectively capture the dynamic nature of news on an ongoing and evolving basis. 

Existing studies such as dynamic propagation-based methods (e.g., Section~\ref{Sec.dynamic-graph-based-methods}), tend to treat samples (i.e., news articles) in a fake news dataset as independent and identically distributed. They formulate fake news detection as an inductive classification task. Each news sample in a dataset is considered individually and classified only based on  information from that sample, e.g., its text content and propagation pattern. In the real world, news items posted on the same topic, in a short time frame, or by closely related users, are often correlated. Such correlations are missed if the news items are treated separately, e.g., transductive methods consider the distribution of all samples in a dataset and how they may be used to address veracity issues. Using such methods, samples distributed closely, e.g., posted by users from the same community, are more likely to be classified into the same class. 

Some heterogeneous social context-based methods~\cite{nguyen2020fang,mehta2022tackling} are based on transductive learning, which have a limited capability to incorporate  contextual  information from related news samples. However, a critical limitation of these methods is that only a limited amount of information is utilised in their models. As a result, these models only focus on a small scope of context, e.g., a subset of users who share news on similar topics. A larger scope including external context, i.e., where the news to be classified is created and posted is still largely opaque, e.g., recent public opinion or attention beyond the immediate social media context. 

To incorporate a larger external context, NEP~\cite{sheng2022zoom} collects a large volume of  news from verified outlets (e.g., Huffington Post, NPR, and Daily Mail) to form the external context.  When classifying a given news item, the verified news articles posted during the period around the time when the target news item was posted are compared with the 
target news item. NEP is not a graph-based method, however, and it is based on external knowledge retrieval and text comparison. A graph-based method considering the external context and temporal information offers a promising future direction. One challenge here is that the external context is typically not as dynamic as social media news, e.g., official news articles are typically published after all information sources are checked and verified.


\subsection{Efficiency and Cost}

The computational complexity and scalability of detection algorithms have not been fully  explored. Fake news detection involves analysing large volumes of textual and multimedia data, which can be computationally intensive.  Scaling fake news detection methods to handle the vast amount of information generated on social media and online platforms can thus be challenging. Systems must be able to process a high volume of news articles, tweets, posts, and multimedia content in real-time.
The processing power and resources required for real-time or large-scale analysis is a major challenge, especially for resource-constrained systems. As more and more graph-based algorithms are proposed with more complicated graph modelling methods, the complexity and scalability demands require further research exploration. 

Secondly, as mentioned earlier, the existing datasets tend to be outdated given the dynamic news topics and  especially the impact of emerging AI techniques. Nowadays, AI has the ability to synthesise text, images and video with very low cost. However, gathering reliable and diverse training data for fake news detection can be time-consuming and costly. The data collection process may involve manual annotation, fact-checking, and verification, which can require human expertise and considerable effort. From a graph-based detection perspective, it would also be necessary to monitor the broader social context, while this is challenging especially for emerging patterns in social media platforms. This also requires access to and use of huge volume of data. 

If any solutions were to be used to tackle the current news veracity issues, then no doubt fake news producers would intentionally craft content to bypass such detection algorithms. This might be through adversarial attacks aiming to manipulate the features or propagation of news articles to deceive detection models. Defending against such attacks requires continuous monitoring, updating models, and staying ahead of evolving fake news techniques. Currently only a few efforts have explored adversarial learning technology \cite{sun2022rumor} and social propagation edge enhancement \cite{mehta2022tackling}.  

Last but not the least, deploying and maintaining an automatic fake news detection system requires appropriate infrastructure and ongoing maintenance. This includes hosting resources, storage, monitoring, and periodic updates to adapt to changing news patterns and techniques. This should potentially be the remit of social media platform providers, but they may argue that others should take this responsibility, e.g., governments. Major platform providers such as Twitter are explicitly offering themselves as the open platform offering a voice for all in the global town square. Many of the most followed accounts belong to individuals who have been demonstrably shown to post factually incorrect materials, hence it is not in the platforms business interests to block such users and/or stop such content from arising.






\bibliographystyle{IEEEtran}

\bibliography{main}

\end{document}